\documentclass[12pt]{article}\usepackage[hyperfootnotes=false]{hyperref}
   \usepackage{epsfig}
      \usepackage{amsmath}
  \usepackage{graphicx}
  \newlength{\abstractwidth}
  \renewcommand{\thefootnote}{\fnsymbol{footnote}}
  \renewcommand{\thanks}[1]{\footnote{#1}} 
  \newcommand{\starttext}{
  \setcounter{footnote}{0}
  \renewcommand{\thefootnote}{\arabic{footnote}}}
  \renewcommand{\theequation}{\thesection.\arabic{equation}}
  \newcommand{\be}{\begin{equation}}
  \newcommand{\bea}{\begin{eqnarray}}
  \newcommand{\eea}{\end{eqnarray}}
  \newcommand{\beq}{\begin{equation}}
  \newcommand{\ee}{\end{equation}}
  \newcommand{\eeq}{\end{equation}}

  \def\ba{\begin{eqnarray}}
  \def\ea{\end{eqnarray}}

  \def\12{{1 \over 2}}
  \def\eq{&=&}

  \def\d{\partial}

  \def\cc{cosmological constant }
  
  \def\simleq{\; \raise0.3ex\hbox{$<$\kern-0.75em
      \raise-1.1ex\hbox{$\sim$}}\; }
   \def\simgeq{\; \raise0.3ex\hbox{$>$\kern-0.75em
      \raise-1.1ex\hbox{$\sim$}}\; }

\def\sig{$\Sigma$}

\def\cdl{Coleman De Luccia}

\def\ba{\bf{a}}

  \def\h3{{\cal{H}}_3}

\def\o3{\Omega_3}

\def\O2{\Omega_2}
\def\o{\omega}
 \def\d2{$dS_{2+1}$}
\def\21{$(2+1)$-dimensional}
\def\31{$(3+1)$-dimensional}
 \def\bi{\begin{itemize}}
  \def\ei{\end{itemize}}

\begin{document}

  \renewcommand{\theequation}{\thesection.\arabic{equation}}
  \begin{titlepage}
  \rightline{SU-ITP-12/11}
  \bigskip

  \bigskip\bigskip\bigskip\bigskip

    \centerline{\Large \bf {The Three Faces of a Fixed Point}}
    \bigskip

  \bigskip \bigskip

  \bigskip\bigskip
  \bigskip\bigskip

  \begin{center}
  {{Daniel Harlow, Stephen H. Shenker, Douglas Stanford, Leonard Susskind }}
  \bigskip

\bigskip
Stanford Institute for Theoretical Physics and  Department of Physics, Stanford University\\
Stanford, CA 94305-4060, USA \\

\vspace{2cm}
  \end{center}

  \bigskip\bigskip

 \bigskip\bigskip
  \begin{abstract}


It has been argued that the only  mathematically precise quantum descriptions of gravitating systems are from vantage points  which allow an unbounded amount of information to be gathered.  For an eternally inflating universe that means a hat, i.e., the asymptotic future of a flat FRW universe. The boundary of the hat (the place where it enters the bulk geometry) is the seat of the FRW/CFT duality.  In this paper we discuss the perturbative and non-perturbative fixed points of FRW/CFT as seen from the three regions which share this boundary.

Perturbatively,  there is nothing universal about the FRW duality; there is a separate construction for each possible initial vacuum.  We explain how bubble collisions induce a non-perturbative flow to a universal fixed point which contains information about the entire landscape.  We also argue that there is a duality between the landscape of de Sitter vacua and a discretuum of very low-dimension relevant operators in the FRW/CFT spectrum.  In principle this provides us with a precise definition of unstable de Sitter vacua.

Along the way we show that BPS domain walls play a special role in
reinforcing the ``persistence of memory" and breaking the symmetries of
the hat.

 \medskip
  \noindent
  \end{abstract}

  \end{titlepage}

  \starttext \baselineskip=17.63pt \setcounter{footnote}{0}


\tableofcontents

\setcounter{equation}{0}
\section{Introduction}

Conventional bulk gravitational degrees of freedom are  useful for
many purposes, but they are   inadequate for  others. They
clearly fail  for certain questions involving quantum
information in the presence of event horizons. The origin of black hole entropy is only one example.
Alternative formulations, in so far as they are known, are holographic
\cite{'tHooft:1993gx}.

It is not obvious whether any of the problems of cosmology are so subtle as
to require the precision of a holographic description. True, inflation and
accelerated expansion imply the existence of event horizons, and primordial
density fluctuations appear to be born out of quantum fluctuations. But
successful descriptions of these phenomena seem to be straightforward enough
that the subtleties of the holographic principle, and of horizon microstates,
appear to be irrelevant. So one might reasonably ask: Who needs a microscopic
holographic description?

The answer may be that the multiverse does. The problem of putting a
probability measure \cite{Linde:1994gy} on the space of vacua has so far defied current formulations of eternal inflation. It's
not that promising phenomenological candidates don't exist---they do---but the
principles underlying them do not.

Many questions suggest themselves. Among them:
\bi
\item  Is the global multiverse\footnote{In this context the term multiverse
could stand for any geometry or cosmology that contains causally disconnected
regions.} described by a single wave function?

    \item Does probability theory need revision in view of the paradoxes of
    exponentially increasing populations \cite{Guth:2011ie}.

\item  Does quantum mechanics make sense on scales so large that no observer
can ever collect data and confirm statistical predictions? A simple example
would be the correlation function between vacua at causally disconnected
points \cite{Bousso:2011up}.

\item Are causally disconnected  degrees of freedom independent,
entangled, or are they complementary descriptions of the same thing?

\item What is the precise meaning of a metastable vacuum? In scattering
theory a metastable state is a pole in the complex energy plane. What is the
mathematical framework in which metastable  de Sitter space is defined?

    \item Is the measure problem so delicate and subtle that resolving it
    requires understanding the above items?

\ei

It seems clear that some new principle is needed to understand these questions.  One conjecture, based on its validity in AdS and Minkowski examples, is that the only  precise representation of the
multiverse involves a holographic  platform from which an observer's past
light-cone encompasses an infinite entropy bound \cite{Bousso:1999xy,Susskind:2007pv,Harlow:2010my}.\footnote{This conjecture also fits well with ideas about the interpretation of quantum mechanics expressed in \cite{Bousso:2011up}.}  Infinite information
capacity would be necessary to encode the entire multiverse or even a small
comoving portion of it. The only candidates in an eternally inflating
multiverse seem to be observers that enter stable supersymmetric bubbles which approach $\Lambda=0$ at late times.\footnote{We expect supersymmetry to be needed for the precise cancellation that sets $\Lambda$ to zero.  The only exception to this, pointed out be Eva Silverstein, seems to be bubbles where the coupling asymptotically goes to zero.}  Such observers are called ``census takers'' \cite{Freivogel:2006xu,Susskind:2007pv,Sekino:2009kv}, and they seem to posess the capability to process arbitrarily large amounts of information.\footnote{A subtlety here is that any ``actual'' census taker has physical limitations that allow only a finite amount of information to be received and stored, but the point is that there is no limit on how large this amount can be.  Moreover nothing in this paper really depends on the existence of actual census takers, below we define them simply as time-like curves.  We thank Edward Witten for many discussions of these issues.}  According to this line of thinking, the key to information-theoretic puzzles (possibly including the measure problem)
is first understanding how to organize the observations of a census taker into a well-defined mathematical theory, and second in clarifying how the information about the larger multiverse is encoded in these observations.

The perturbative observations of a census taker have been studied in a series of papers \cite{Freivogel:2006xu,Park:2008ki,Sekino:2009kv,Park:2011ty}, which have established a local conformal structure on the census-taker's cosmic sphere.  This program is far from complete, but the results so far are consistent with a conjectural ``FRW/CFT'' duality that realizes a census taker's observations as correlation functions in a CFT coupled to gravity living on his/her cosmic sphere.\footnote{The name ``FRW/CFT'' turns out to be somewhat inaccurate, in that non-perturbative effects in the multiverse will break the conformal symmetry.  This is somewhat analogous to the term ``AdS/CFT'', which has since been widely applied to systems that are not conformal.  In our case a more accurate name might be ``eternal inflation/fractal flow'' duality, using a term we will introduce below, but this is rather tedious so we will continue to use ``FRW/CFT'' with the understanding that it stands for a more general set of ideas.}  These perturbative calculations have a rather unappealing feature in that they seem to depend on the particular vacuum the census taker ends up in and also on the vacuum he/she nucleated from.  Moroever they have so far not been able to shed light on how to extract information about the larger multiverse.  In this paper we will claim that once non-perturbative bubble nucleation and thermal fluctuation are taken into account, there is a universal fixed point describing the late-time observations of all census takers and encoding information about the entire landscape.\footnote{More precisely we will say that each connected moduli space of $\Lambda=0$ vacua has a single type of census taker with a unique fixed point.  If there are disconnected moduli spaces then it seems likely that each has its own fixed point.  The case of moduli spaces that are connected only at infinity is subtle, we comment briefly on it below.}  This argument has been made before \cite{Susskind:2007pv,Sekino:2009kv}, but we are now able to be much more specific about it.

We will also focus on the existence and importance BPS domain walls, connecting the FRW region and a bubble of negative cosmological constant \cite{Freivogel Horowitz Shenker}. If such domain walls exist they significantly modify both the boundary theory and the evolution within the FRW bulk geometry. We will see that such domain walls ``shred'' the census taker's sky and turn it into a fractal of measure zero, breaking the conformal symmetry to an unusual kind of scale invariance. This breaking is a form of ``persistence of memory'', which although it sounds threatening, has the positive effect of selecting a preferred frame of reference and eliminating the ambiguity in the choice of time-variable.

In making our arguments, we will make constant use of a partition of the multiverse into three separate regions.  The three regions are defined by their causal relationship to the census taker, and the interplay between them is a new crucial part of the FRW/CFT duality.  We refer to the complementary descriptions of the three regions within the dual theory as the three faces of the fixed point.

\section{Three Regions}
We begin by giving general definitions of some of the standard FRW/CFT concepts.\footnote{These geometric definitions are really just semiclassical notions, but the idea is that they will enable a bulk discussion that suggests a precise boundary version.}
\begin{itemize}
\item A ``census taker'' is a timelike curve of infinite future extent, whose nearby geometry approaches a piece of Minkowski space at late times.
\end{itemize}
Given a particular census taker, we can divide up spacetime into three regions based on their causal relation to that census taker.
\begin{itemize}
\item The causal past of the census taker, meaning all points which can influence the census taker's wordline by sending in massless particles on null geodesics, we denote as $\mathcal{C}$.
\item A spacetime point is in ``region III'' if it is not in $\mathcal{C}$.
\item A spacetime point in $\mathcal{C}$ which can influence some point in region III is in ``region II''.
\item A spacetime point in $\mathcal{C}$ which is not in region II is in ``region I''.
\end{itemize}
In \cite{Freivogel:2006xu} \cite{Sekino:2009kv} the emphasis was on the region I, but in this paper we will be interested in all three regions.  We can similarly partition the future boundary as follows.
\begin{itemize}
\item The census taker's ``hat'' is the future boundary of region I.
\item The ``rim'' of the hat, which we denote $\Sigma$, is the future boundary of region II.
\item The ``fractal future'' is the future boundary of region III.
\end{itemize}
Clearly $\Sigma$ is a special part of the boundary, it is the interface between the census taker's observations and the fractal future.  $\Sigma$ is the seat of the FRW/CFT duality.

These definitions can be applied to any spacetime that has a census taker; for example in empty Minkowski space only region I exists, while for an eternal black hole in Minkowski space region III is behind the horizon and the rest of the space is region II.  For an evaporating black hole all three regions exist.
\subsection{The Coleman De Luccia Geometry}
\begin{figure}[h]
\begin{center}
\includegraphics[scale=1]{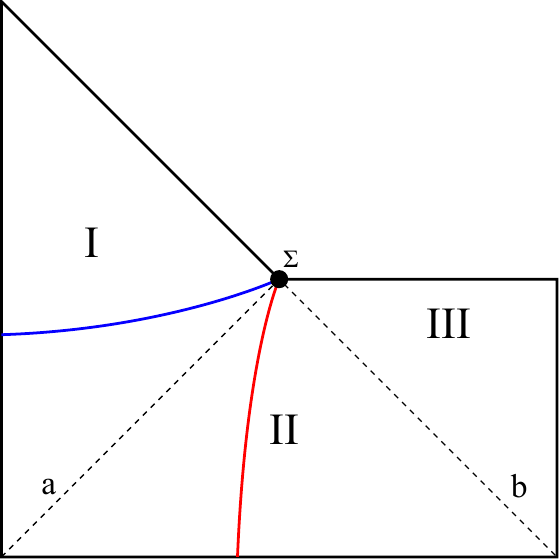}
\caption{The various regions of the Lorentzian \cdl \ geometry for the decay of dS space to $\Lambda=0$.  They are separated by the dashed lines \textbf{a} and \textbf{b}, and the colored lines are representative slices that are preserved by the symmetry.  So the blue line is an FRW slice in region I and the red line is a $dS_3$ slice of the domain wall.  The census taker can be taken as the timelike geodesic on the left side of the diagram.  The geometry has a lower piece that is the time reverse of the upper half, but only this upper half is relevant for tunelling.}
\label{fig:1}
\end{center}
\end{figure}

The intuitive meaning of the three regions in the multiverse is best illustrated by considering the example of the Coleman De Luccia spacetime describing a bubble of Minkowski vacuum expanding into an ancestor de Sitter space \cite{Coleman:1980aw}.  We can think of the Minkowski bubble as a point on the supersymmetric moduli space of flat vacua, in which case the moduli fields will be massive in the de Sitter region of the geometry.  The Penrose diagram of this geometry is depicted in Figure
\ref{fig:1}, with the three regions labelled. We have also called the lightlike boundary between region I and region II ``surface \textbf{a}'' and the lightlike boundary between region II and region III ''surface \textbf{b}''.  Surface \textbf{b} is the horizon of the census taker.

We see that region I is the open FRW region in which the census taker finds itself at late times, that region II is the domain wall connecting the two vacua, and that region III is the ``rest of the multiverse''.  $\Sigma$ has the topology of a two-sphere, and it can be thought of as the late-time sky of the census taker.

This geometry has a large degree of symmetry, which manifests itself differently in the three regions.  In region two we can choose coordinates of the form
\be
ds^2=a(X)^2\left(dX^2-d\omega^2+\cosh^2\omega d\Omega_2^2\right),
\ee
which clearly has \d2 symmetry, that is $O(3,1)$, acting along slices of fixed $X$.  $X$ runs from $-\infty$ to $\infty$.  This metric can be thought of as a warped product of \d2 over a line, which suggests the possibility of a dimensional reduction \cite{Karch:2003em,Alishahiha:2004md} to a genuine \d2 theory.  The future boundary of this lower dimensional theory is an $S^2$ which can again be thought of as $\Sigma$, and indeed it will be very convenient below for us to think of FRW/CFT as the theory at future infinity of this \d2.

\begin{figure}[h]
\begin{center}
\includegraphics[scale=.4]{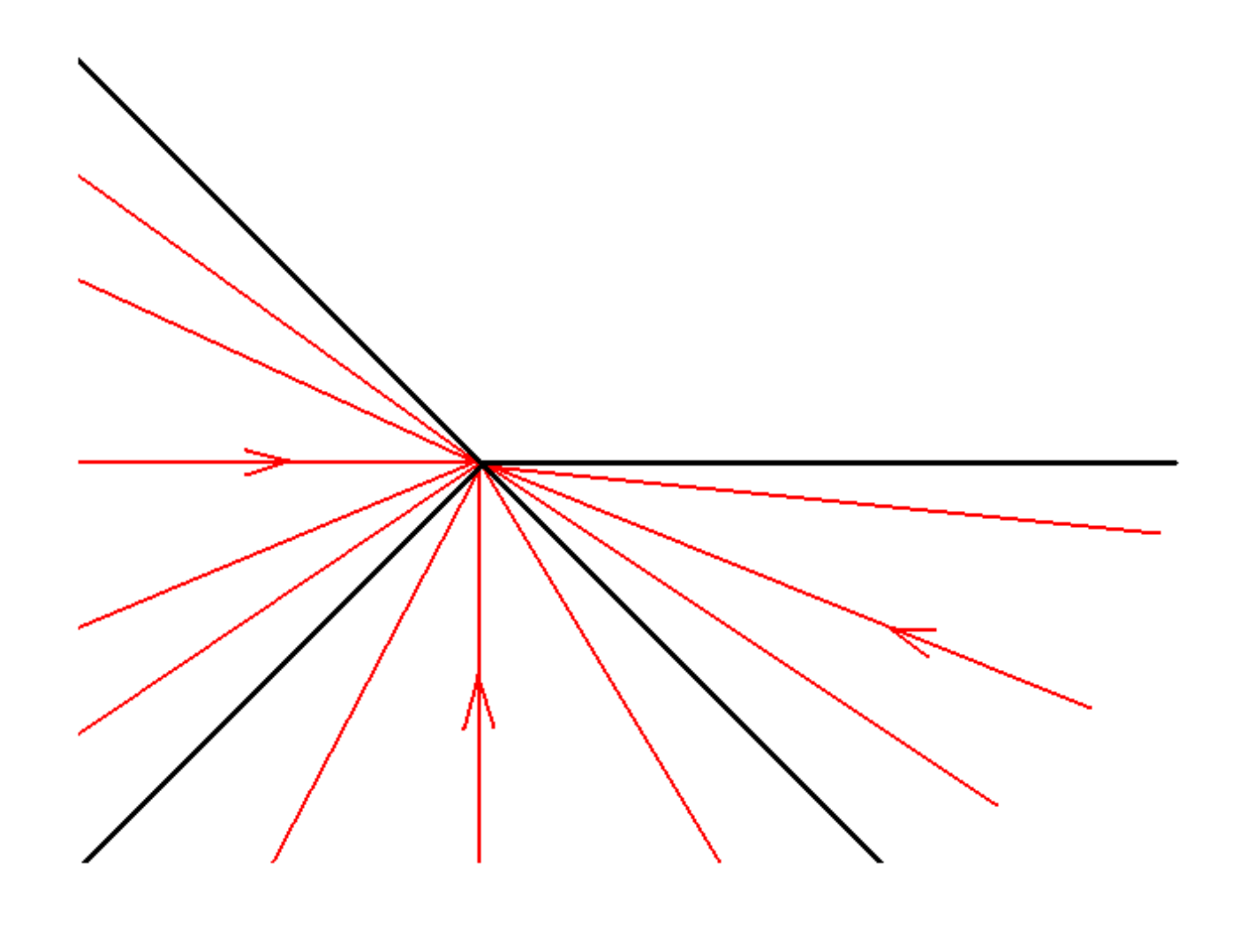}
\caption{Orbits of $O(3,1.)$}
\label{fig:3}
\end{center}
\end{figure}

Similarly in region I and region III we have coordinates of the form
\be\label{frw-metric}
ds^2=a_{\small{I},\small{III}}(T)^2 \left(-dT^2+d\mathcal{H}_3^2\right),
\ee
where $\h3$ is the hyperbolic three-plane
\be
d\h3^2 = dR^2 + \sinh^2 R \ d\Omega_2^2,
\label{hyper plane}
\ee
and $T$ runs from $-\infty$ to $\infty$ in region I and from $-\infty$ to $0$ in region III.  Both regions again have $O(3,1)$ symmetry; in region I this is the FRW symmetry of the census taker.  Among other things it suggests at least perturbatively that different census takers in region I should see similar physics.  In fact we will see in section \ref{terminalsect} that nonperturbative effects break this symmetry, but it is still a very useful organizing principle.  The symmetry orbits near $\Sigma$ are shown in figure \ref{fig:3}.  They all converge at $\Sigma$.

In region III the scale factor at late times ($ T\to 0^-$) behaves as
\be
a_{III}(T)\sim -\frac{1}{HT},
\ee
where $H$ is the hubble constant in the ancestor $dS_{3+1}$ vacuum.

In region I the scale factor has the asymptotic form
\be
a_I(T) \to {1\over H}e^T
\label{FRW scale factor}
\ee
for $T\to \pm\infty$.  To keep the discussion as simple as possible we will sometimes use the thin-wall approximation and set $a_I(T)$ equal to ${1\over H}e^T$

We emphasize that in general $\Sigma$ has a number of meanings:
\bi
\item
First of all $\Sigma$ is the spatial boundary of region I.  It is the location of the holographic degrees of freedom of
the census taker's FRW region.
\item  $\Sigma$ is  the  spacelike future boundary of region II, or what we
called \d2.
\item Finally  $\Sigma$ is the boundary of region
III---the-rest-of-the-multiverse, beyond the census taker's horizon.
\ei
\subsection{New Coordinates}
 We are mainly interested in the immediate vicinity of the surface  $\Sigma.
 $ There is of course an infinite volume of space-time in all three regions
 in the vicinity of any point of $\Sigma,$ so we really don't lose anything
 by zooming in on such a point. From the FRW point of view we are looking at
 the asymptotic behavior of the geometry out along a particular direction.
For this purpose we
introduce light-cone coordinates in region I,
\be
T^{\pm} = T \pm R,
\label{light cone coords}
\ee
and take the limit of large $R$. We also introduce two flat tangent space
coordinates, $Y^1, \ Y^2$ to describe the neighborhood of a point on the two
sphere:
\be
d\Omega_2^2 \to 4 dY^idY^i
\ee
The limiting FRW metric \ref{frw-metric} satisfies
\be
H^2 ds^2 = -e^{(T^+ + T^-)}dT^+ dT^- + e^{2T^+}dY^idY^i
\ee

In order to attach the FRW patch to regions II and III it is helpful to
change variables,
\bea
U^+ \eq -e^{-T^+} \cr \cr
U^- \eq e^{T^-}
\label{U=e^T}
\eea
yielding the metric
\be
H^2 ds^2 = \left({1\over U^+}\right)^2 [-dU^+ dU^- + dY^idY^i]
\label{U-flat}
\ee
The range of $U$ corresponding to region I is
\bea
U^+ &<& 0  \cr
U^- &>& 0
\eea

The metric \ref{U-flat} can be smoothly attached to the bulk 4-dimensional de
Sitter space (regions II and III). The metric for de Sitter space in
conformal coordinates satisfies
\be
H^2 ds^2 = \left({1\over U^0}\right)^2[-dU^+ dU^- + dY^idY^i]
\label{ds U}
\ee
where $U^0 = (U^+ + U^-)/2.$

The metrics \ref{U-flat} and \ref{ds U} agree along the surface $U^- =0$.
Thus, to sew the metrics together we extend $U^-$ to negative values. In
region $U^-<0$ the range of variation of the coordinates is $U^+ + U^- <0.$

The procedure outlined above corresponds to a light-like domain wall at $U^-
=0.$ The generalization to a time-like domain wall is simple.
\bea
H^2ds^2 \eq {-dU^+ dU^- + dY^2 \over (U^0)^2} \ \ \ \ \ (U^-< qU^+)  \cr \cr
H^2ds^2 \eq {-dU^+ dU^- + dY^2 \over 4(1+q)^2 (U^+)^2} \ \ \ \ \  (U^->
qU^+)
\label{zoom metric}
\eea
where the domain wall is given by $U^- = q U^+.$ The parameter $q$ is
positive and less than one.

The three regions I, II, and III, are given by:
\bea
U^+ &<& 0, \ \ \ \ \ U^- > 0 \ \ \ \ \ \ \ \ \ \ \ \ \ \ \ \  \rm{Region \ I}
\cr
U^+ &<& 0, \ \ \ \ \ U^- < 0 \ \ \ \ \ \  \ \ \ \ \ \ \ \ \ \  \rm{Region \
II} \cr
U^+ &>& 0, \ \ \ \ \ U^0 < 0 \ \ \ \ \ \ \ \ \ \ \ \ \ \ \ \ \rm{Region \
III}
\eea
Regions I and II are clearly visible to the census taker while region III is
behind the horizon.

Note that in this formulation, the orbits of $O(3,1) $ are planes of the
form
$$\frac{U^+ }{U^-}= c $$
with $c$ being a constant.
 The census taker's horizon separating regions II and III is  defined by the
 surface,
\be
U^+ = 0
\ee

None of the above depends in any important way on the thin wall
approximation.
There is a simple generalization that applies to all \cdl \ geometries beyond
the thin-wall limit. The general form involves a single function
$F\left(\frac{U^+}{U^-}\right)$
\be
H^2ds^2 = F\left(\frac{U^+}{U^-}\right) \left[ { -dU^+dU^- + dY^2 \over
(U^+)^2          } \right]
\label{general metric}
\ee

The important feature of this metric is its scale invariance under the
rescaling of all coordinates,
\bea
U^{\pm} &\to& \lambda U^{\pm} \cr
Y^i  &\to& \lambda Y^i
\label{scaling}
\eea
and the $E(2)$ symmetry under translations and rotations of $Y.$ In addition
there are other transformations, which together fill out the conformal
transformations on $\Sigma.$
What we are really doing by focusing on the vicinity of a point is the analog
of going to Poincare coordinates in anti de Sitter space.

Note that the domain wall, being an orbit of  $O(3,1),$ corresponds to a plane
at constant $\frac{U^+}{U^-}.$

\setcounter{equation}{0}
\section{The Perturbative Fixed Point}
\subsection{Initial Conditions}

\begin{figure}[h]
\begin{center}
\includegraphics[scale=.4]{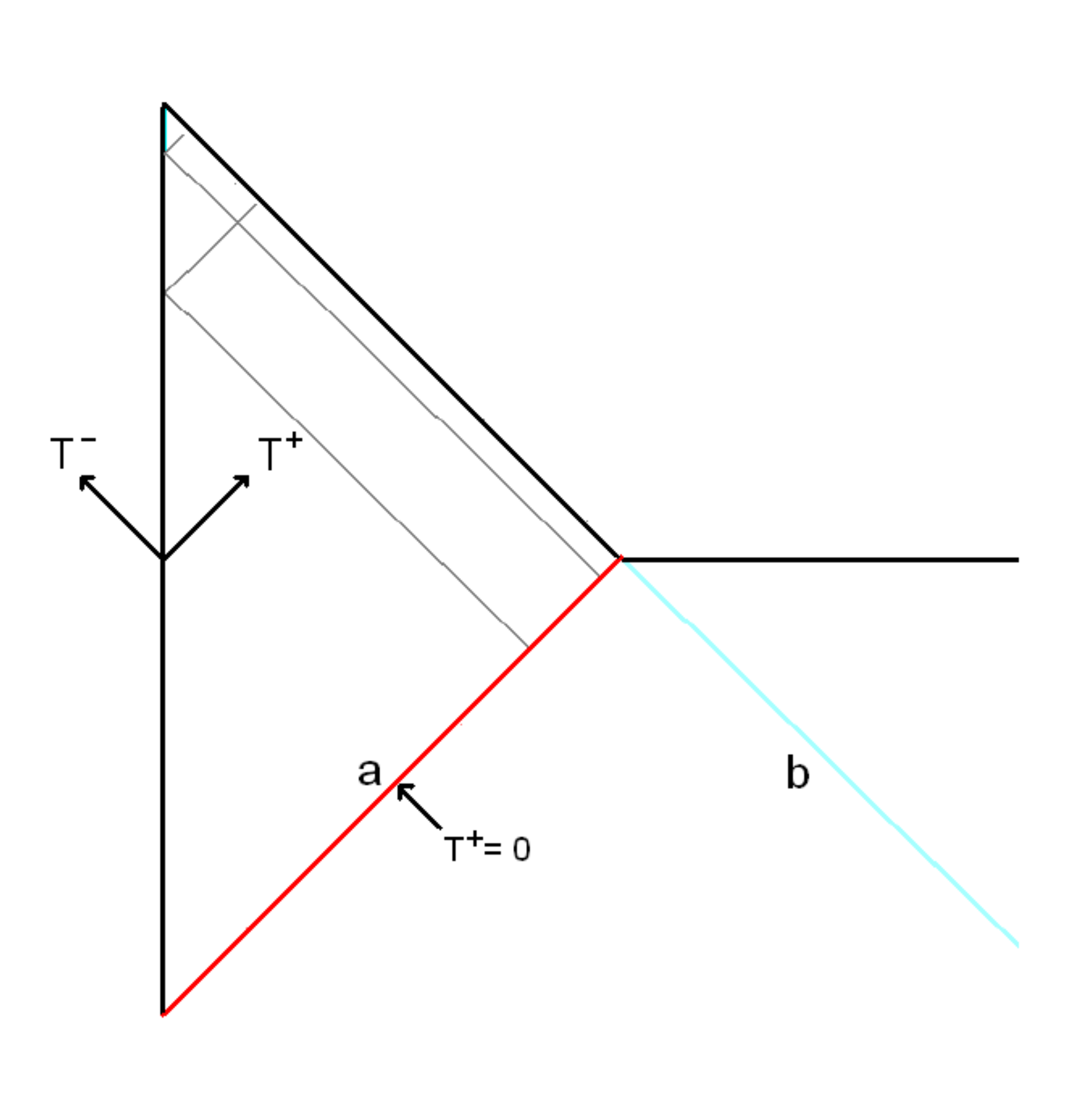}
\caption{Initial conditions shown in red. Radiation entering at $T^+ $
is deposited on the hat at $T^-.$}
\label{fig:4}
\end{center}
\end{figure}

The initial conditions\footnote{Initial conditions for the FRW patch should
be distinguished from global initial conditions for the multiverse. } for the
FRW universe in region I are naturally formulated on the hypersurface
$a(T)=0$ shown as the red line in Figure \ref{fig:4}.
From equations \ref{frw-metric}, \ref{hyper plane}, and \ref{FRW scale
factor}
one sees that this hypersurface is  given by  $T^-=-\infty$ or equivalently
$U^-=0.$
Points on the initial condition surface have coordinates $T^+, Y^i.$
Classically, initial conditions can be formulated in terms of the values of
various fields $\phi(T^+, Y^i)$ on $T^-=-\infty.$ These  fields  propagate
into the FRW region and eventually deposit their information on the hat at
$T^+=\infty.$  Quantum mechanically, the initial data can be formulated in
terms of correlation functions of fields on the surface $\bf{a}.$ \
An  example would be the two point function of a scalar field, evaluated  on
$\bf{a}:$
\be
\langle \chi(T_1^+, Y_1^i) \ \  \chi(T_2^+, Y_2^i)  \rangle
\label{<XX>}
\ee

The perturbative study of such correlation functions was undertaken in
\cite{Freivogel:2006xu} \cite{Sekino:2009kv} \cite{Park:2011ty}  where the
emphasis was on extrapolation to $\Sigma$ from region I.
We will briefly state the results of these papers for the case of the
two-point function of a minimally coupled scalar field.
In the leading semiclassical approximation the 2-point function \ref{<XX>} in
region I is a sum of three terms, $G_1,
 \ G_2$ and $G_{flat}$

These terms were  explained in \cite{Sekino:2009kv} in equations (5.14),
(5.20), and (5.21).
Only one term, $G_2$ survives the limit $T^-\to -\infty.$ It has the form of
a sum over CFT correlation functions,
\be
G_2=\sum_{\Delta}\tilde{G}_{\Delta} e^{-\Delta(T_1^+ + T_2^+)}(1-\cos
\alpha)^{-\Delta}
\label{G-2}
\ee
where the $\Delta$ are a set of scaling dimensions which are either real or
which come in complex conjugate pairs. The real parts of the $\Delta$  are all
positive. The $\tilde{G}_{\Delta}$ are a set of numerical coefficients, and
$\alpha$ is the angular
distance between points $1$ and $2$ on $\Omega_2. $ Adapting the formula to
the coordinates $U^{\pm},Y^i,$
\be
G_2=\sum_{\Delta}\tilde{G}_{\Delta}
|U_1^+|^{\Delta} |U_2^+|^{\Delta}
| Y_1-Y_2  |^{-\Delta}
\label{flat G2}
\ee

Each contribution has the form of a scalar-scalar correlation function in a
2-dimensional CFT. The factors $|U_1^+|^{\Delta}=e^{T^+}$ are interpreted as the
wavefunction renormalization factors with $U^+$ being the renormalization
length scale. Obviously the RG-flow toward the ultraviolet is represented by
the evolution $T^+\to \infty.$ The set of scaling dimensions characterize the
fixed point on $\Sigma.$

 The fields on $\bf{a}$ are the initial conditions for
 radiation entering the FRW region.  The overwhelming majority of the emitted
 photons (and of the information that they carry) comes in at asymptotically
 large $T^+.$ It is therefore controlled by the behavior of the fixed point.
  Obviously, the semiclassical initial condition and the resulting radiation
  is sensitive to the properties of the ancestor vacuum and the domain wall,
  but little else. In section \ref{terminalsect} we will show that the situation is very
  different when non-perturbative effects are included.

\subsection{Dimensional Reduction in Region II}\label{ds2sect}

Consider the geometry of any surface of fixed $U^+/U^-.$ Such surfaces are
orbits of $O(3,1) $ and are therefore homogeneous spaces. In regions I and
III, where the orbits are space-like, the fixed $U^+/U^-$ surfaces are
Euclidean hyperbolic planes. In region II the orbits are time-like and  are
$2+1$-dimensional de Sitter spaces. In this region the ratio $U^+/U^-$
defines a compact space-like coordinate. The geometry has the form of a
warped product of $2+1$-dimensional de Sitter space and a line interval.  By
the change of variables
\bea
u_0 \eq -\sqrt{U^+U^-} \cr \cr
e^{2X} \eq \frac{U^-}{U^+}
\eea
the metric \ref{general metric} in region II can be brought to the form
\be
H^2 ds^2=a(X)^2     \left(    \frac{-du_0^2 + dY^2}{u_0^2} + {dX^2} \right)
\label{warped ds}
\ee
which manifestly has the form of a warped product of $(2+1)$-dimensional de
Sitter space (in flat slicing and conformal coordinates) and the $X$-line.
Although $X$ runs from $-\infty$ to $+\infty$ the $X$-line is of finite
length since $a(X)$ tends to $e^{-|X|}$ at $X=\pm \infty.$

Note that by defining
\be
u_0=-e^{-\omega}
\ee
the metric \ref{warped ds} can be put in the familiar exponentially expanding
form
\be
H^2 ds^2=a(X)^2     \left(   - d\omega^2 + e^{2\omega}dY^2 + {dX^2} \right).
\label{exp warped ds}
\ee

For future reference we note that
\be
\omega \pm X = T^{\pm}
\label{w+-X=T+-}
\ee
and the metric \ref{exp warped ds} can be written in the form,
\be
H^2 ds^2=a(X)^2
\left(  -dT^+dT^- + e^{(T^++T^-)}dY^2  \right).
\ee

Returning to the metric as written in \ref{exp warped ds}, the radius of curvature of each leaf of the foliation (each value of $X$) is given by
$\frac{a(X)}{H}$ and therefore vanishes at the endpoints. This formally
gives rise to a continuum of Kaluza Klein modes labeled by a continuous
momentum-like quantum number  $k.$ The Kaluza Klein mass (in units of $H$) is
given by
\be
\mu^2 = 1+k^2
\label{2+1 mass}
\ee

 The geometry in \ref{warped ds} and \ref{exp warped ds} will be called \d2.
Because the $X$-interval is compact the graviton wavefunction is normalizable
which implies that there is a genuine $(2+1)$-dimensional dynamical
gravitational degree of freedom \cite{Alishahiha:2004md} in \d2.
The domain wall of the \cdl \ geometry is a surface of constant $X.$ One can
roughly think of \d2 \ as being the geometry of a thick domain wall and the
$(2+1)$ dimensional gravitational field as being bound to the domain wall.

A causal patch of \d2 \ is defined as the interior of the backward lightcone
of a point on $\Sigma. $  Since \d2 is a dynamical de Sitter space its causal
patches have  entropy. The entropy given by  standard compactification
arguments is
 \be
 S_{2+1} = \frac{m_p^2}{H^2}\int a(X)^2 \ dX
 \label{entropy of d2}
 \ee
 where $m_p$ is the four-dimensional Planck mass.

Causal patches of \d2 \ also have a temperature which is generally of order
$H$.
Thermal radiation emitted from the census taker's horizon (surface $\bf{b}$
in Figure \ref{fig:1}) is seen as radiation entering the FRW region,
eventually being deposited on the hat.

The most interesting thing about \d2 \ from the present perspective is that
its future boundary is $\Sigma.$ Thus one may speculate that there is a
second duality involving $\Sigma;$ namely a $(2+1)$-dimensional dS/CFT
duality.

There is one feature of \d2 \ that makes it conceptually different from
ordinary stable de Sitter space. The far future of de Sitter space consists
of many causally disconnected causal patches, which cannot communicate either
with each other,  or with any third party. Moreover, from experience with black holes we have
learned that degrees of freedom on the two sides of a horizon are not independent.

For these reasons one may seriously
doubt whether there is operational meaning of  a global quantum description of
de Sitter space. For example products of fields at separated comoving points
cannot be measured by any physical observer.

One answer to this objection is that if de Sitter space eventually  decays to
flat space by reheating, then the entire space becomes visible and large
scale correlations are measurable.
This may be  true in the context of conventional inflation, but it is irrelevant to eternal inflation, which by definition
never globally reheats.

The lower dimensional \d2 \ is very different in this respect. Although observers who remain in region II fall out of causal contact with
one another, and therefore cannot communicate, they can all be seen by the
census taker. Correlations between distant points in \d2 \ do have
operational meaning, as would a wavefunction. One could say that the Census
Taker is the ``Meta-observer" that overlooks the entire global \d2.

\subsection{Correlations in Region II}

The FRW/CFT correspondence of \cite{Freivogel:2006xu} and
\cite{Sekino:2009kv} is  one of three different limits, all converging on
$\Sigma$ from the three regions, I, II, and III. All three  are interesting:

\bi
\item Region I contains a hat and hats are the only places in which an
infinite amount of information can be stored and observed.

\item Region II extends back into the causal past of the hat. It contains the
entire history of the census taker's causal patch, possibly including
multiple bubble nucleations. We are undoubtedly in the causal past of some
hat, and therefore in a region II.

\item As for region III, ultimately we either want a description of the
rest-of-the-multiverse or to know why one is not possible.

\ei

Thus it is extremely interesting that these three regions come together and
share the same boundary, namely $\Sigma.$ One would  like to understand as
much as possible about the theory living on $\Sigma$ from the three different
complementary perspectives.
Earlier work concentrating on region I was largely perturbative. In this
section we extend the perturbative theory to region II, mostly as a warmup
for the non-perturbative analysis in  Section 8. The geometry of region II is
\d2;  it is natural to suppose that it is described by some version of dS/CFT
\cite{Strominger:2001pn}, \cite{Maldacena:2002vr}. To explore and confirm
this possibility we will analyze how correlation functions behave as their
arguments tend to $\Sigma.$
Consider a $(3+1)$-dimensional minimally coupled scalar field $\chi$
propagating in region II. In region I $\chi$ may be massless,  but without
supersymmetry to protect it, $\chi$  will be massive in region II. More
precisely  $\chi$ will have an $X$-dependent mass $m(X).$

Late-time correlation functions in $(d+1)$-dimensional de Sitter space have a
form which resembles those of a Euclidean $d$-dimensional CFT
\cite{Strominger:2001pn}. We expect that the same is true for \d2 \ in the
limit that the arguments  approach $\Sigma.$ The method of calculation  in
region II can be borrowed directly from \cite{Freivogel:2006xu} and was
briefly described in \cite{Harlow:2010my}.

 In $(d+1)$-dimensional dS/CFT the spectrum of operator dimensions $\Delta$
 is discrete
  and connected with the  masses of the bulk fields. For
  minimally-coupled-scalar fields
the scaling dimensions are given in terms of the mass by
\be
\Delta =\frac{d}{2} - {\sqrt{\frac{d}{4}^2-\frac{m^2}{H^2}}}.
\label{ds dimensions}
\ee
The $\Delta$ are real for sufficiently small mass, but for large mass they
become complex with a real part equal to $1.$

For \d2 \ \ref{ds dimensions} becomes
\be
\Delta =1 - {\sqrt{1-\frac{\mu^2}{H^2}}}.
\label{d2 dimensions}
\ee
where $\mu$ is the appropriate mass given by \ref{2+1 mass}.
To calculate the correlation functions in \d2 \ we expand the four
dimensional fields in a Kaluza Klein expansion. Consider for example, a
massless minimally coupled field in the ancestor vacuum.
The wave equation for the KK modes $u(X)$ has the form of a one-dimensional
time-independent Schrodinger equation,
\be
[-\partial^2_X + V(x)]u(X) =k^2u(x)
\ee
where the potential $V$ is given by
\be
V(X) = \frac{a''(X)}{a(X)}-1.
\ee
Since $a(X) \to e^{-|X|}$ the potential goes rapidly to zero at large $|X.|$
It is also easily seen to be attractive with a single normalizable bound
state with $k^2 = -1.$ This corresponds to a $(2+1)$-dimensional  mass
$\mu^2=0.$

If the $(3+1)$-dimensional  mass of the scalar is not zero in the ancestor
vacuum, the potential changes by amount $\delta V= m^2 a(X)^2$ and the
eigenvalue shifts from
$k^2 = -1.$ For small values of $m$  ($m<H$) the eigenvalue remains real and
satisfies
\be
k= \pm i(1-\mu^2)
\ee
where $\mu^2$ is of order $\frac{m^2}{H^2}.$ We will see that this
corresponds to an operator dimension of magnitude $\mu^2$ at the fixed
point.

All the other modes are in the continuum and oscillate. The corresponding
 dimensions are
\be
\Delta =1 \pm ik.
\ee

Complex dimensions are not surprising in the present context \cite{Strominger:2001pn}
but
a continuum of dimensions is disturbing, and not what we expect from a CFT.
However, this situation is familiar from \cite{Freivogel:2006xu} and has an
elegant resolution. The integrals over $k$ that define correlation are
contour integrals that can be deformed so that the correlation functions
become discrete sums of terms. We will repeat the argument shortly.

For the moment let us consider the bound and continuum eigenvectors of $[-\partial^2_X
+ V(x)].$ The eigenvectors consist of the bound state wavefunction
$$u_B(X)$$ and a set of incoming scattering states $u_l(x)$ and $u_r(x)$
corresponding to waves coming in from the left $(x<<0)$ and right
$(x>>0.)$ Asymptotically these waves have the form
\bea%
u_l(k,X) \eq e^{ikx}+R_l(k) e^{-ikx}  \ \ \ \ \ \ (x<<0) \cr \cr
u_l(k,x) \eq T_l(k)e^{ikx} \ \ \ \ \ \ \ \ \ \ \ \ \ \ \ \ \  (x>>0)
\label{u-left}%
\eea%
\bea%
 u_r(k,X) \eq e^{-ikx}+R_r(k) e^{ikx}  \ \ \ \ \ \
(x>>0) \cr \cr u_r(k,x) \eq T_r(k)e^{-ikx} \ \ \ \ \ \ \ \ \ \ \ \ \
\ \ \ \ (x<<0)%
\label{u-right}
\eea%
 where in both cases $k$ is positive.

 One can easily prove the following:

The wave functions $u_l(k,X) $ and $ u_r(k,X)$ are analytic
 functions of $k.$ On the real axis the continuation to negative $k$
 is given by
 \be
u_{l,r}(-k,X) = u_{l,r}^{\ast}(k)%
\label{u-cont star}
 \ee
It follows that%
\bea%
R_{l,r}(-k)\eq R_{l,r}^{\ast}(k) \cr \cr %
T_{l,r}(-k)\eq T_{l,r}^{\ast}(k)%
\label{RT cont star}%
\eea%
\bea%
T_r(k) \eq T_l(k) \cr \cr %
R_r(k)\eq -R_l(k) \frac{T(k)}{T*(k)}
\label{left to right}%
\eea%

More generally for complex $k,$%

\bea%
T_r(k) \eq T_l(k) \cr \cr %
R_r(k)\eq -R_l(k) \frac{T(k)}{T(-k)}%
\label{RT cont}%
\eea%
The
behavior of the wave functions is more complex for finite $x$ but
fortunately we know a good deal about them.

Let us consider the two-point function of the scalar field $\chi$
\be%
  \left\langle  \chi(X_1, \omega_1)
\chi(X_2,
\omega_2)    \right\rangle %
\ee %
in
the limit $\omega_{1,2} \to \infty.$

Define $\alpha$ to be the angular distance on $\Omega_2$  between the two points $1,2,$
and  $z=\cos{\alpha}.$
For $\omega \to \infty$
The correlation function can be written as a sum over the eigenvectors
$u_B,$
$u_l,$ and $u_r.$

\bea
&& \langle  \chi(X_1, \omega_1)
\chi(X_2,
\omega_2)    \rangle \cr \cr
\eq \frac{1}{4\pi i a(X_1) a(X_2)}\cr \cr
&&
\int_0^{\infty}\frac{dk}{\sinh{\pi  k}}
\left[
            u_l(k,X_1)  u^\ast_l(k,X_2)+u_r(k,X_1)  u^\ast_r(k,X_2)
\right]
e^{-(1+ik)\bar{\omega}}(1-z)^{-(1+ik)} \cr \cr
&-& \frac{1}{4\pi i a(X_1) a(X_2)} \cr \cr
&& \int_{0}^{\infty}\frac{dk}{\sinh{\pi  k}}
\left[
          u_l(k,X_1)  u^\ast_l(k,X_2)+u_r(k,X_1)  u^\ast_r(k,X_2)
\right]
e^{-(1-ik)\bar{\omega}}(1-z)^{-(1-ik)}  \cr \cr
&+& \rm{bound \ \ state \ \ contribution.}
\label{pot}
\eea

where $\bar{\omega} = \omega_1 +\omega_2.$

Now we use the fact that on the real $k$ axis the wave functions can be
analytically continued to negative $k,$
\be
u_{l,r}(-k, X)=u^{\ast}_{l,r}(k, X),
\ee
to write the continuum contribution to the correlation function as a contour
integral over the real axis:

\bea
&& \langle  \chi(X_1, \omega_1)
\chi(X_2,
\omega_2)    \rangle \cr \cr
\eq \frac{1}{8\pi i a(X_1) a(X_2)}\cr \cr
&&
\int_{-\infty}^{\infty}\frac{dk}{\sinh{\pi  k}}
\left[
            u_l(k,X_1)  u_l(-k,X_2)+u_r(k,X_1)  u_r(-k,X_2)
\right]
e^{-(1+ik)\bar{\omega}}(1-z)^{-(1+ik)} \cr \cr
&-& \frac{1}{8\pi i a(X_1) a(X_2)} \cr \cr
&& \int_{-\infty}^{\infty}\frac{dk}{\sinh{\pi  k}}
\left[
          u_l(k,X_1)  u_l(-k,X_2)+u_r(k,X_1)  u_r(-k,X_2)
\right]
e^{-(1-ik)\bar{\omega}}(1-z)^{-(1-ik)}  \cr \cr
&+& \rm{bound \ \ state \ \ contribution.}
\label{contour}
\eea

In this form the correlation function is a  sum over a continuum of complex
conformal dimensions. This is so because the correlation function for a
scalar field of scaling dimension $\Delta$ has the form
\be
G_{\Delta}\sim (1-z)^{-\Delta}.
\ee

To turn the expression in \ref{contour} into a discrete sum of terms we use
the fact that the wave functions $u(k,X)$ are analytic functions of $k$
except for discrete poles. The poles in the upper half plane are associated
with bound states and are strictly on the imaginary $k$ axis. In the lower
half plane they occur on the imaginary axis, or they occur in pairs symmetrically
located with respect to the imaginary  axis. The integrands also have poles at the
imaginary integers due to the factors $\sinh^{-1}(\pi k).$

For large $\omega$ the contour for the first integral,
\be
 \frac{1}{8\pi i a(X_1) a(X_2)}
\int \frac{dk}{\sinh{\pi  k}}
\left[
            u_l(k,X_1)  u_l(-k,X_2)+u_r(k,X_1)  u_r(-k,X_2)
\right]
e^{-(1+ik)\bar{\omega}}(1-z)^{-(1+ik)},
\ee
can be closed in the lower half plane and one picks up a discrete set of
terms of definite dimension.
The other integral is closed in the upper half plane and also produces a
similar set of discrete terms. Each term is accompanied by a residue function
that depends on $X.$ One may think of these as  Kaluza Klein wave functions, but they
do not form an orthonormal family.

Let us consider a specific contribution from the first integral.  The
contribution from a pole at $k=i\kappa$ is,
\be
F(X_1,X_2)e^{-(1-\kappa)\bar{\omega}}\frac{1}{(1-z)^{1-\kappa}}
\ee
For this term $\kappa <0$ since the contour is closed in the lower half
plane. This means that the dimension $\Delta = (1-\kappa)$ is greater than
$1.$ There are no very low dimension correlation functions from this term.

The same is true of the other integral which gives contributions
\be
F(X_1,X_2)e^{-(1+\kappa)\bar{\omega}}\frac{1}{(1-z)^{1+\kappa}}
\ee
but this time with $k>0.$ We find the continuum contributions are of the
conformal kind but with dimensions all satisfying $\Delta >1.$

Later we will be especially interested in very low dimensional contributions.
The only place that they can come from, perturbatively, is the bound state
contribution. In the (unphysical)  massless case  the bound state is a
solution of the Schrodinger equation with eigenvalue $k^2 = -1.$    In the
small mass  case the eigenvalue gets shifted to a value $k=\pm i(1-\epsilon)$
where
\be
\epsilon \sim \frac{m}{H}
\ee
In that event the bound state contribution has the form
\be
\langle  \chi(X_1, \omega_1)
\chi(X_2,
\omega_2)    \rangle_{BS}=\frac{ u_B(X_1)}{a(X_1)}\frac{u_B(X_2)}{a(X_2)} e^{-\epsilon (\omega_1+
\omega_2)}\frac{1}{(1-z)^{\epsilon}}
\ee

For completely massless scalars the bound state wave function $U_B(X)$ is exactly equal
to $a(X)$ and the ratios $\frac{ u_B(X)}{a(X)}$ are unity. For $m \neq 0$ they do not cancel but asymptotically for large $|X|$ the ratios satisfy
\be
\frac{ u_B(X)}{a(X)}\to e^{\epsilon X}.
\ee

In other words it is the correlation function for a dimension $\epsilon$
scalar on the surface $\Sigma.$ The approach to the surface is controlled by
the wave function renormalization factors $e^{-\epsilon (\omega_1+
\omega_2)}.$

\subsection{Back to the Initial Condition Surface }

The initial conditions for the census taker's bubble is given by the values of
correlation functions on the
 surface $\bf{a}.$ That surface is
 defined by allowing $X\to - \infty$ and $\omega \to \infty$
with $\omega+X$ fixed, or from  \ref{w+-X=T+-}, with $T^+ $ fixed.

The easiest way to take that limit is to go
back to equation \ref{contour} and take the limit using \ref{u-left}, \ref{u-right}, and \ref{left to right}. In the limit $X\to - \infty, \ \  \omega \to \infty$ one finds

\bea
&& 8\pi i \left\langle  \chi(X_1, \omega_1)
\chi(X_2,
\omega_2)    \right\rangle \cr \cr
\eq e^{-{X_1}}e^{-{X_1}}
\int_{-\infty}^{\infty}\frac{dk}{\sinh{\pi  k}}
\left[
2\cosh{k(X_1-X_2)} +R(k)e^{ik(X_1+X_2)}+ R(-k)e^{-ik(X_1+X_2)}
\right]  \cr \cr
&& \left\{ e^{-(1+ik)(\omega_1+\omega_2)}(1-z)^{-(1+ik)}
-
e^{-(1-ik)(\omega_1+\omega_2)}(1-z)^{-(1-ik)}
\right \} \cr \cr
&& + \rm{bound \ \ state \ \ contribution.}
\label{contour2}
\eea
The factors $e^{-{X_1}}e^{-{X_1}}$ are the asymptotic values of $$\frac{1}{a(X_1)} $$ and $$\frac{1}{a(X_2)} $$ relating the correlation function to $\hat{G}.$

The bound state contribution is extremely simple.
\bea
8\pi i \langle  \chi(X_1, \omega_1)
\chi(X_2,
\omega_2)   \rangle
\eq  e^{-\epsilon (\omega_1+X_1)}e^{-\epsilon (\omega_2+X_2)}
\frac{1}{(1-z)^{\epsilon}} \cr \cr
\eq
e^{-\epsilon T_1^+}e^{-\epsilon T_2^+}
\frac{1}{(1-z)^{\epsilon}}
\eea
The remaining contributions have the same form with $\epsilon$ being replaced by
operator dimensions $\Delta.$ We see that the correlation function on the initial condition surface has the form in Equation \ref{G-2}.

The structure of these correlations emphasizes the deep
relationship between RG flow of the
$(2+1)$-dimensional dS/CFT to the $T^+$ dependence of the initial condition
on the surface $\bf{a}.$ Since the surfaces of constant $T^+$ are the Census
Taker's past light-cones, we see how the RG flow is intimately connected with
the evolution of the census taker's observations.

\setcounter{equation}{0}

\section{Doubled dS/CFT in Region II}

Whether or not a global CFT description of ordinary de Sitter space makes
operational sense, the corresponding dS/CFT description of \d2 seems
well-motivated.
Since $\Sigma $ is the asymptotic future of \d2 \ the dS/CFT would be a
description of this surface. However it would not be the same limit as the
FRW/CFT theory describes: the two limits approach  $\Sigma$
from different directions. Approaching $\Sigma$ along spacelike surfaces in
region I is very similar to approaching the boundary of Euclidean AdS space.
In region II  $\Sigma$ is approached along timelike directions. There is a
big difference in how observables and expectation values are defined in the
two cases \cite{Harlow Stanford}. Nevertheless the two theories of $\Sigma $
must be intimately connected.

This section is not essential to the arguments of the present paper but it is
included for a couple of reasons. The first is just to show how earlier
conjectures about $dS/CFT$ can be applied to \d2. The second is that it may
suggest an eventual  microscopic framework.

Applying the arguments of  \cite{Maldacena:2002vr}  we assume that the
Wheeler DeWitt wavefunction of \d2 \ is given by the partition function of a
Euclidean two-dimensional CFT with fields $A$ and sources $h.$ The sources
are the usual \31 \ fields. Among the them are the components of the spatial
metric, light scalars, and other bulk fields. They also include the
``multiverse fields " \cite{Harlow:2011az} that we will explain later. The
$A$ on the other hand, are analogous to gauge fields and other boundary
degrees of freedom in ADS/CFT  duality. We will refer to them as gauge
fields.

We assume the action for the gauge theory is
\be
I=\int d^2Y L(A;h).
\ee
The Wheeler DeWitt wavefunction is a functional of the sources,
\be
\Psi(h)=\int e^{\int d^2Y L(A;h)} dA
\ee

 The gauge fields $A$ are not the observables that we are interested in. Our
 interest lies in the statistics of the sources. The probability for a given
 configuration of sources is given by the Born rule,
 \be
 P(h) = \Psi^\ast(h)\Psi(h)
 \ee
which can be directly computed using a doubling trick,
 \be
 P(h) = \Psi(h)=\int e^{\int d^2Y \{L(A;h)+L^\ast(B;h)\}} dAdB
 \label{double}
 \ee
 where $B$ are a second set of gauge fields. Note that the sources are the
 same in the $A$ and $B$ terms.

 Now let's consider the expectation values of functionals of $h.$ Let $F(h)$
 be such a functional.
 \be
 \langle  F(h) \rangle = \int F(h) e^{\int d^2Y \{L(A;h)+L^\ast(B;h)\}}
 dAdBdh
 \ee
 In other words the expectation values of bulk fields at late time is given
 in terms of a \it doubled \rm theory with dynamical fields that include two
 copies of the gauge fields and a single copy of the sources.
The fact that we have to integrate over the sources means that the machinery
for computing expectation values  is a form of two-dimensional Euclidean
quantum gravity. If one gauge-fixes  to  conformal gauge, then the theory is
described as two copies of  ``matter" coupled to a Liouville degree of
freedom.

One might expect that the Liouville field defined in this way would be the
same Liouville field that appeared in the FRW/CFT theory, but that would be a
mistake;  the geometry of $\Sigma$ is defined in two different ways which
are not continuations of each other. In the dS/CFT description of \d2 the
gravitational field is an average over the $X$ coordinate. The continuation
of $X$ into region I is FRW time. The Liouville field in FRW/CFT is not an
average over time; it is time.

From another more technical point of view the two Liouville theories cannot
be the same. It is known that the logarithm of the Wheeler DeWitt wave
functional does have a Liouville contribution in $(2+1)$ dimensions
\cite{Banks}.
As in the FRW/CFT case the magnitude of the central charge is of order the
ancestor entropy. However, in the \d2 \ case the central charge is imaginary
whereas in the FRW case in is real and negative.

In fact the difference between the theories becomes even clearer in the
doubled theory. In that case the central charge of the dS/CFT Liouville
theory is zero, since it cancels when computing
$\Psi^\ast \Psi.$ By contrast the large negative central charge of the
FRW/CFT puts the theory into the deep semiclassical regime with a very stable
spherical saddle point. It is evident that in order to stabilize the spatial
geometry in $(2+1)$-dimensional de Sitter space one must fix a time variable
in some way. Despite the fact that the Liouville fields are different in the
dS/CFT and FRW/CFT descriptions of $\Sigma$ there must be a close connection
between the theories.

It is apparent that the limit $u^0\to 0$ defines the UV fixed point behavior
of the doubled theory in much the same way that $U^+ \to 0$ defined the
fixed point of the FRW/CFT theory. In fact we expect the operator dimensions
to be the same.

The doubled theory is a machine for calculating correlations in \d2. Some
approximation to it should be able to reproduce the correlation functions of
the last section including the low dimension operators associated with very
light scalars. Such low dimension correlations fall off slowly and obviously
control the largest scale features. One might think there are not very many
low dimension operators. However this is not so. In the next section we will
explain how non-perturbative effects create an enormous discretuum of
extremely low dimension fields on $\Sigma,$ which essentially describe the
entire landscape of metastable vacua. These low dimensional fields may
provide a rigorous classification of the landscape.

The existence of low dimension scalars means that the fixed point of the
doubled CFT is unstable with respect to flowing toward the infrared. Even in
the perturbative case, if the mass of the scalar field $\chi$ is smaller than
$H$ there is a relevant operator in the doubled CFT.
How do we understand these instabilities\footnote{Whether to think of the UV
or the IR as the initial state of an RG flow is ambiguous. In condensed
matter physics one usually thinks of the starting point, or initial
condition, being the UV end, the output being the IR behavior. In cosmology
the IR corresponds to the initial condition and the UV to the late-time
behavior.}?

The UV limit in region II is late time, i.e., $\omega \to \infty.$ Flowing
toward the infrared means running toward early time. As one runs toward early
time, relevant perturbations will eventually become strong. This means that
things become sensitive to global initial conditions\footnote{We will assume
that the global initial condition consists of specifying an initial de Sitter
vacuum with no bubbles on an initial global spacelike  hypersurface.  }
\cite{Garriga:2006hw}.
The closer the UV is to the fixed point, the longer (into the past) it will
take to feel the relevant operators. To put it another way, very small
relevant perturbations mean the initial condition is far in the past, while
large relevant
perturbations mean that the bubble nucleated close to the initial condition
surface.

\setcounter{equation}{0}
\section{The Non-Perturbative Fixed Point}
\subsection{Non-Perturbative Effects}\label{NPsect}

Describing eternal inflation in terms of the hat immediately
raises the question of which hat? A \cdl \ instanton describes
tunneling from a particular ancestor to a particular final flat vacuum. Are
there a large class of different descriptions for each possible initial and
final state? Or are they all the same? The problem is exacerbated by two
facts. The first is that the number of classically stable de Sitter vacua is
likely to be extremely large. The second is that they may all be subject to
non-perturbative instabilities and therefore not be accurately defined.

The answer that we will propose is that the FRW/CFT duality is not about any
specific transition. Mobility on the landscape, due to non-perturbative
effects, cause an RG flow to a common final
 UV fixed point that  contains information about the entire landscape.

The non-perturbative effect that drive this RG flow is the nucleation of  \cdl
\ bubbles. Such  bubbles can collide with the Census
Taker's bubble and perturb the initial condition on the surface $\bf{a}.$  To the census taker these collisions appear as
disc-like disturbances concentrated over a small region of the $Y$
coordinates. The later the colliding bubble nucleates the smaller the
disturbed region. From the point of view of the FRW/CFT theory living on \sig \ this type of
configuration might be thought of as a 2-dimensional Euclidean instanton.

There are several types of bubble collisions to consider:

 \begin{figure}[h]
\begin{center}
\includegraphics[scale=.7]{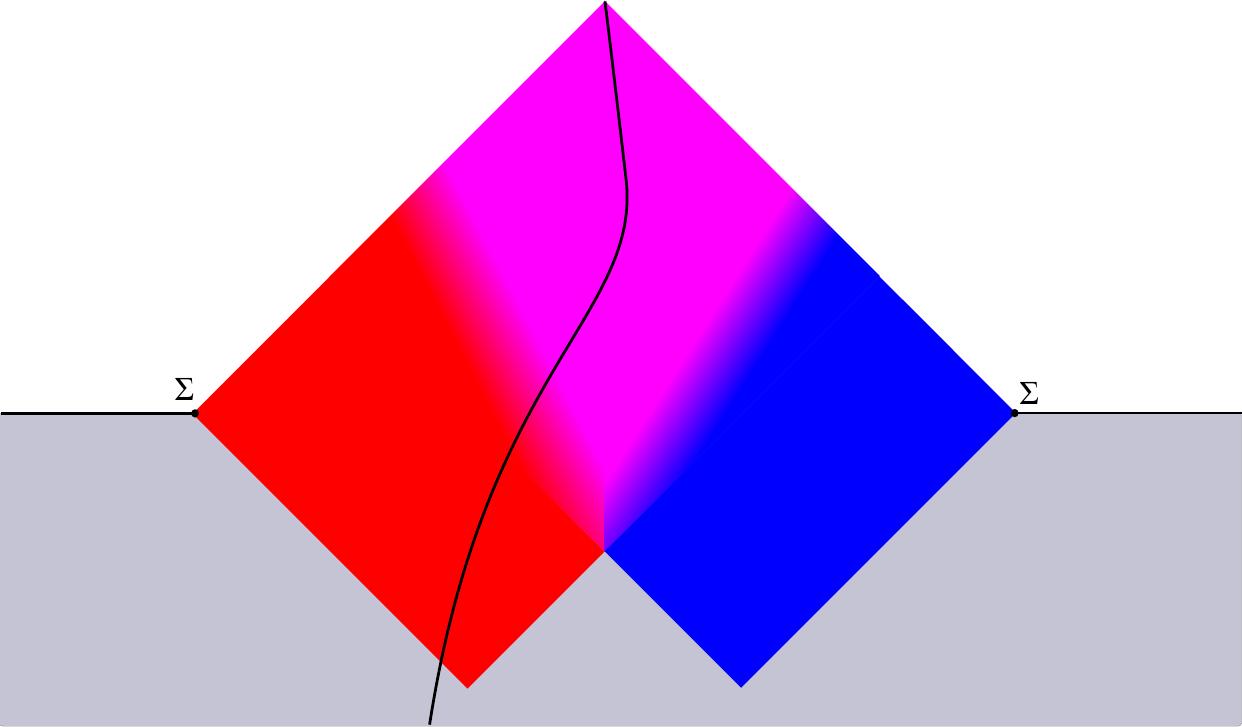}
\caption{A slice of the collision of two $\Lambda=0$ bubbles on the same moduli space. A census taker that enters from either side ends up in the same hat.  The two indicated points are part of the boundary two-sphere $\Sigma$.}
\label{fig:5}
\end{center}
\end{figure}
\bi

\item Collision with another FRW bubble on the same moduli space: This is shown in  Figure
\ref{fig:5}.
No domain wall is created by the collision although some energy is released
and radiated into
the FRW regions, and the two bubbles merge to form a single hat.  The vacuum in the
colliding bubble may or may not be at the same point in the moduli space as the
original bubble. If it is not, there is still no domain wall but the two
vacua will bleed into each other. What this means is that such collisions
will contribute to the fluctuations of the massless moduli fields in the FRW
region, and also on $\Sigma.$ 
Note in Figure \ref{fig:5} that some part of what would have
been in region II if only the red bubble nucleated now becomes part of region I; we can think of collisions with other bubbles on the same moduli space as ``pushing out'' $\Sigma$ over some angular region.  This is much like the
displacement of a black hole horizon by an in-falling shell of matter.

\begin{figure}[h]
\begin{center}
\includegraphics[scale=.4]{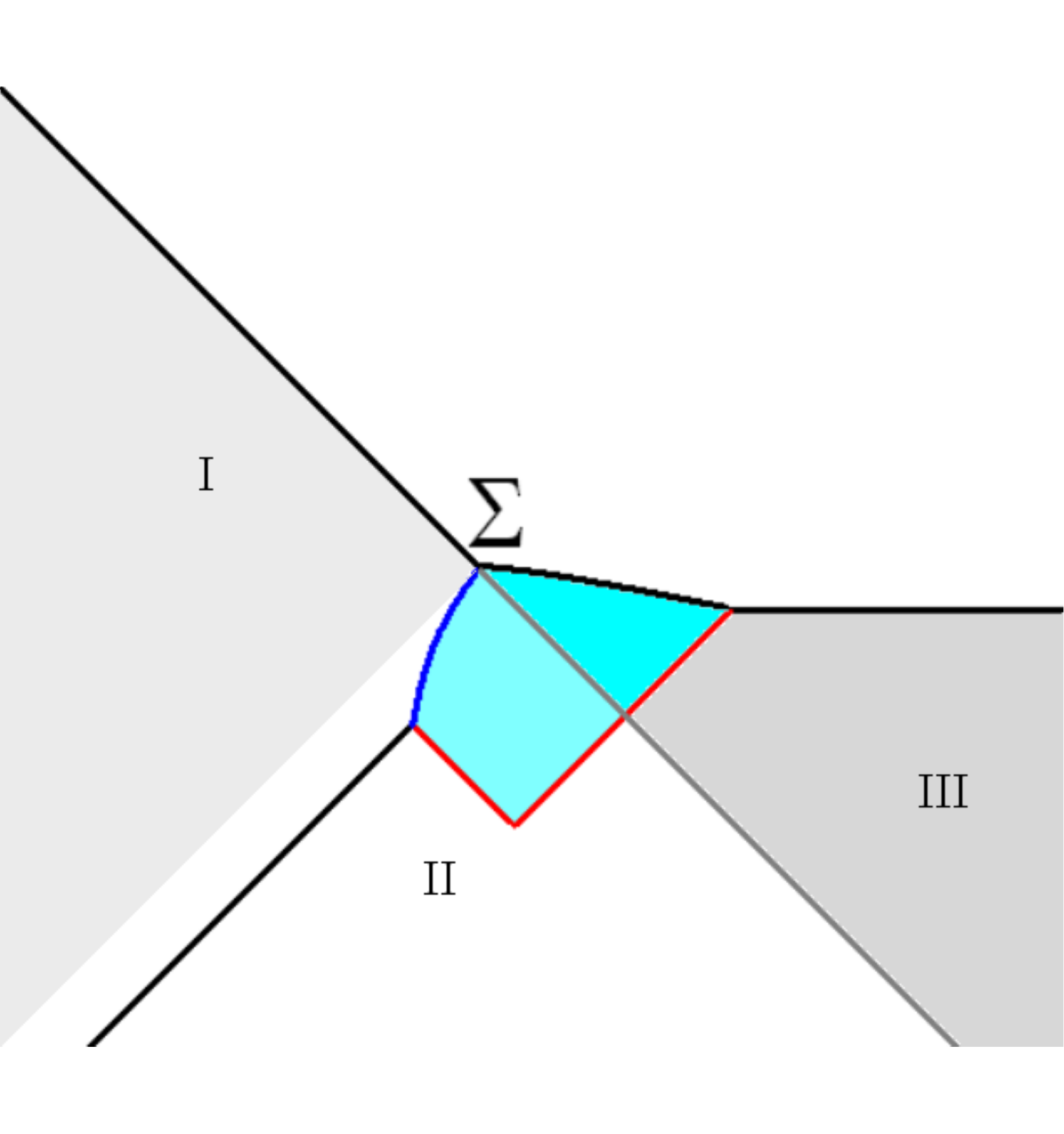}
\caption{Collision with a de Sitter bubble. The light grey and dark grey
regions are in I and III respectively. The white region is in I.}
\label{fig:6}
\end{center}
\end{figure}

\item Collision with a de Sitter bubble: This is shown schematically in
Figure \ref{fig:6}.

The collision results in a domain wall, shown as dark blue, between the
census taker's bubble and the de Sitter bubble. It neither destroys the hat
or $\Sigma$ but it obviously changes the initial conditions for large $T^+$
or equivalently small $|U^+|.$ This can be thought of as ``pulling in'' $\Sigma$ over an angular region.

The future boundary of the colliding bubble is an inflating boundary similar
to that of the ancestor.

\begin{figure}[h]
\begin{center}
\includegraphics[scale=.8]{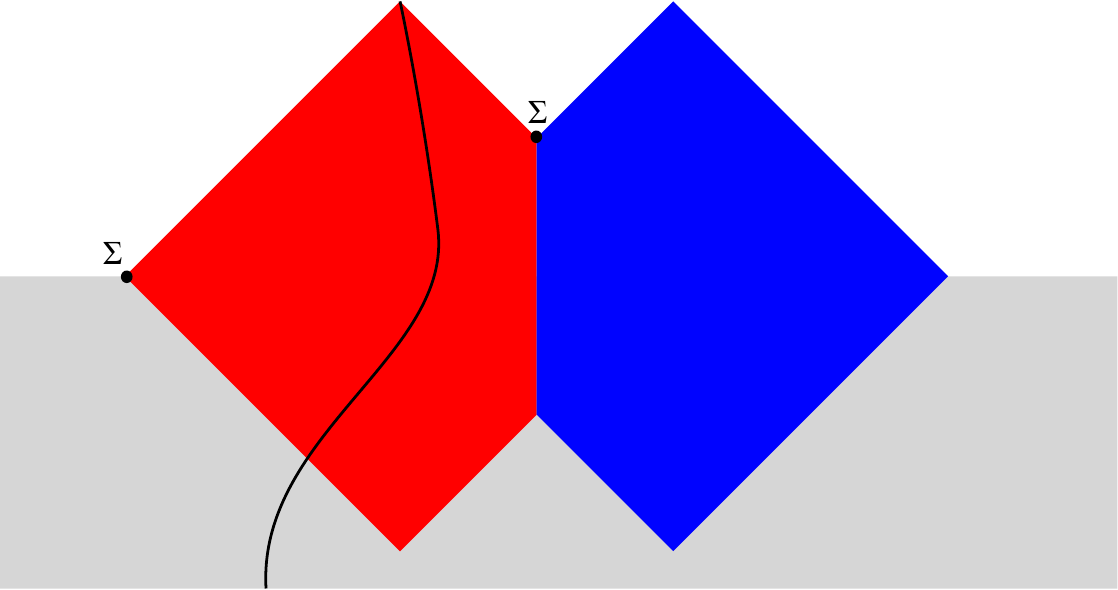}
\caption{Collision of two $\Lambda=0$ bubbles which are on different moduli spaces.  The labelled points are part of $\Sigma$ for the drawn census taker.}
\label{fig:dismod}
\end{center}
\end{figure}

\item  Collision with a flat bubble on a disconnected moduli space. If, as we expect, there are disconnected moduli spaces (for example vacua with different supersymmetry) then a collision between them will produce a domain wall. The geometry of such a collision will tend to the Vilenkin-Ipser-Sikivie  geometry of \cite{Vilenkin:1984hy}. The domain wall will accelerate away from both regions. The intrinsic geometry of the domain wall will be locally de Sitter and will have an acceleration temperature. From the point of view of a census taker on one side or the other, this type of collision is very similar to a collision with a de Sitter bubble.  Unlike the case where both bubbles are on the same moduli space, there are now two hats produced as shown in figure \ref{fig:dismod}.

\item Collision with a crunching vacuum of negative \cc \ which could be a
supersymmetric ADS type vacuum: This is the most dangerous and most poorly
understood case. In general it will lead to a domain wall. The tension of the
domain wall is bounded from below by a BPS bound. There are two cases
\cite{Freivogel Horowitz Shenker}. In the  first case the tension is greater
than the BPS bound, and the domain wall accelerates away from the Census
Taker's bubble. The picture is similar to Figure \ref{fig:5} except that the
future boundary of the bubble is a singularity instead of an inflating
region.  An important fact is that for this case the domain wall is eternally accelerating and
    therefore remains hot during its entire infinite history. This means that
    it is not dead. It is the source of new nucleation events.
             \begin{figure}[h]
\begin{center}
\includegraphics[scale=.4]{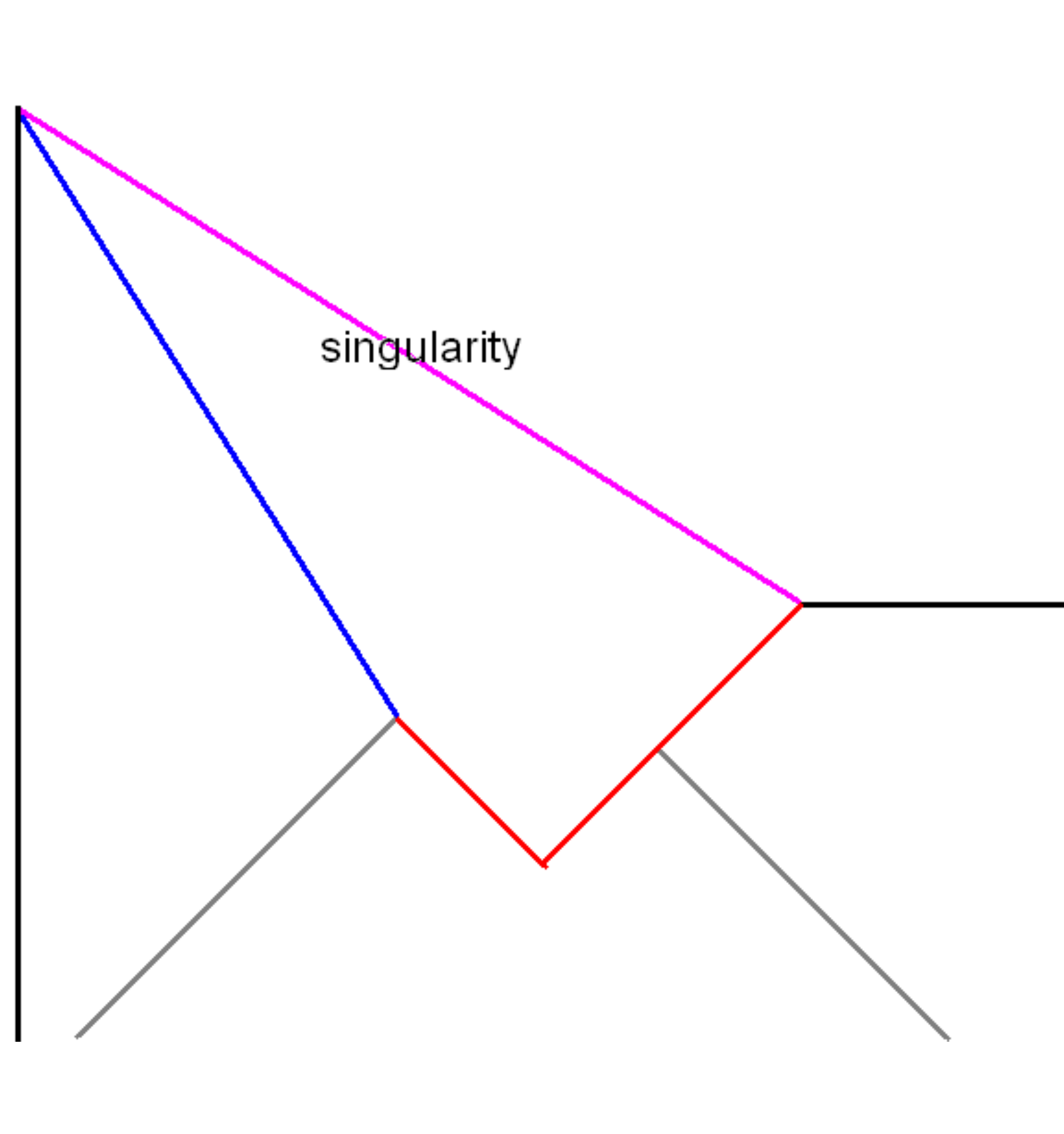}
\caption{Saturating the BPS domain wall bound. The domain wall appears to push $\Sigma$ all the way up to the top of the hat, but remember that this happens only over a small angular region. }
\label{fig:7}
\end{center}
\end{figure}

    \item Alternatively if the domain wall between the crunch and the census taker's
    bubble saturates the BPS bound then it does not accelerate. It penetrates
    the census taker's bubble and eventually hits time-like infinity. This is
    shown in Figure \ref{fig:7}. The conformal diagram makes the collision
    look much more dangerous to the census taker than it really is. For one
    thing the perturbation only occupies a small angle in the census taker's
    sky, and for another, the tip of the hat is really an infinite region,
    most of which is far from the census taker.

\ei

In sections \ref{confsect}-\ref{terminalsect} we will discuss how these nonperturbative effects affect FRW/CFT, but we first build up some intuition from a different source.

\subsection{Some Lessons from Trees}
There is  evidence that perturbative correlation functions
in de Sitter space have conformal limits on the future boundary. Non-perturbatively, when transitions between de Sitter vacua are included,
the evidence is more limited and only applies if there are no terminal vacua. Terminal vacua destroy conformal invariance but leave scale-invariance unbroken. In this section we will review the results of \cite{Harlow:2011az} regarding these issues. For the moment the discussion is general and includes the cases of \31 \ de Sitter space and also \d2.

We begin with a landscape consisting of de Sitter vacua, assuming
for the moment that there are no terminal vacua with either
vanishing or negative vacuum energy. The de Sitter vacua are labeled
by a ``color" label $n.$ To describe the population of colors a
cutoff time must be introduced. Call the cutoff time $t.$ For the
moment the precise definition of time will not be specified. Bubbles
are allowed to nucleate before the cutoff, and then grow and
populate a portion of future infinity. The fraction of  comoving
volume of future infinity populated by color $n$ is called $P(n).$

To regulate the bubble population ignore all bubbles whose
nucleation-centers occur later than $t.$   A bubble that nucleates
earlier than $t$ leaves a colored ball on future infinity. One can
ask how much of the coordinate volume on future infinity  is covered
by color $n$ when the cutoff is $t.$  The answer is governed by rate
equations that were analyzed in
\cite{Garriga:1997ef}\cite{Harlow:2011az}. We will closely follow
the arguments  of \cite{Harlow:2011az} with slight changes of
notation.

One begins with a set of rate equations \cite{Garriga:1997ef}
describing the population as a function of cutoff time $t,$
\be%
\frac{dP_n}{dt} =     \dot{P}_n=\sum_m (\gamma_{nm}P_m
-\gamma_{mn}P_n)
  \label{rate}
\ee
where $\gamma_{nm}$ is the dimensionless transition rate for a
bubble of type $n$ to nucleate in a vacuum of type $m.$ The equation
can be written,%
\be%
  \frac{dP_n}{dt} = G_{mn}P_m
\ee%

If one adds the ingredient of detailed balance the transition rates
satisfy \be \gamma_{nm} = e^{S_n}M_{n m} \ee where $M_{nm}$ is a
symmetric matrix. In this equation $S_n$ represents the
 de Sitter entropy of the $n^{th}$ vacuum.

Making the change of variables \be \Phi_n = e^{-S_n}P_n \ee one
finds the equation \be \dot{\Phi}_n = \sum_m (M_{nm}
e^{\frac{S_m+S_n}{2}} \Phi_m - M_{mn}e^{S_m}\Phi_n). \ee When
written out in matrix notation the equation has the form \be
\dot{\Phi}_n = \sum_m T_{nm} \Phi_m \ee where $T_{nm}$ is symmetric.
Therefore there are a complete set of orthogonal eigenvectors
$\Phi_I(m)$ of $T$ which evolve by simple rescaling, %
\be%
\Phi_I(m, t) = e^{-\Delta_I t} \Phi_I(m, 0)
\ee %
where
$e^{-\Delta_I}$ is the $I^{th}$ eigenvalue of $T_{nm}.$

The $\Delta $ are all positive except for the leading eigenvector
which has $\Delta = 0.$ This means that a generic initial condition
will flow to a stationary probability distribution, and by
inspecting the equations one can see that the stationary
probabilities $P_n$ satisfy
\be
P_n = \frac{1}{Z}e^{S_n}
\label{station}
\ee
 where $Z$ is a normalization factor.

 This stationary probability distribution does not represent any individual local minimum on the landscape. Rather it
 represents a universal bath of
 bubbles at asymptotic time. The distribution as the cutoff tends to
 infinity becomes independent of the initial conditions, and not
 surprisingly  is scale invariant. But the evidence from both
 \cite{Freivogel:2009rf} and the tree-like model
 \cite{Harlow:2011az} is that if detailed balance is
 satisfied then a stronger symmetry prevails, namely
 that the asymptotic population is
 invariant with respect to conformal transformations of future infinity.

In \cite{Harlow:2011az} it was shown that there is a vast discretuum
of
 conformal \it multiverse fields \rm that are associated with the
 eigenvectors
 $\Phi_I.$ One can identify a  multiverse field  with each color $n.$  The
 value of each field is either $0$ or $1$ depending on whether a point on the
 future boundary is that color or not. Multiverse fields have not played a role
 in the dS/CFT context although they certainly should. The sources, or what
 is the same thing, the boundary conditions on the CFT partition function in
 \ref{double} must specify which color vacuum occupies each region of space.

 One may choose the basis of multiverse fields to be labeled by color, or
 alternatively,  superpositions with definite scaling dimensions can be
 defined by the eigenvectors $\Phi_I.$ The scaling dimension of $\Phi_I$ is
 $\Delta_I,$ and there are of course a great many of them if the Landscape is
 rich.

 The values of the $\Delta$ are determined by the rate equations and are
 generally of order the transition rates $\gamma.$ Assuming the existence of
 a great many metastable long lived vacua, there will also be an equally
 large number of very low dimension multiverse fields. The stationary
 eigenvector with dimension zero, \ref{station}, is identified with the unit
 operator in the sense of a CFT. All the other dimensions are positive.

The introduction of terminal vacua significantly modify these conclusions, and in particular, destroy conformal symmetry.
 We will continue to use the color
notation $n$ for de Sitter vacua, and collectively lump the terminal
vacua into the ``the terminal." The effect of the terminal is to
introduce an additional term into the rate equations,
\be
  \frac{dP_n}{dt} =     \sum_m (\gamma_{nm}P_m -\gamma_{mn}P_n)-\gamma_n
  P_n
  \label{ignore terminal}
\ee %
where $\gamma_n$ is the total rate  to decay from color $n$ to the terminal. It is
obvious from these equations that all the $P_m$ tend to zero at late
time. The loss of probability is of course exactly compensated by
the increasing probability $P_t$ for the terminal.

Quantitatively, the
 eigenvalues of \ref{ignore terminal} are all less than $1$ and the
 scaling dimensions $\Delta$ are all positive. The smallest of them,
 $\Delta_D,$
 defines the \it dominant \rm eigenvector. The dominant eigenvector
 replaces the stationary eigenvector.
 In
addition there is a discretuum of very small sub-leading scaling
dimensions $\Delta_I,$ all satisfying $\Delta_I > \Delta_D.$

Eventually the terminal probability tends to unity and all others to
zero.
 The fixed point is replaced by
\bea
P_m \eq 0 \cr P_t \eq 1
\eea

However, the situation is not as dull as it seems. Garriga and  Vilenkin \cite{Garriga:2008ks}
have speculated that there may be a ``conformal field theory on a fractal" living
on the inflating fractal that remains on future infinity after the
terminal is removed. In \cite{Harlow:2011az}  it was shown that an
interesting structure survives in the tree-like model, although
there is no evidence that it is a field theory. The structure is
defined by asking conditional questions of the following kind:

Given a point $p,$ what is the probability that it has color $n$
given that it has not been swallowed up by a terminal bubble?

Similarly, Given two points $p_1$ and $p_2,$ what is the conditional
probability that they have color $n_1$ and $n_2,$ given that neither
has been swallowed up by a terminal bubble?:

Such questions define a collection of conditional correlation
functions
 \bea
 && C_n(p) \cr \cr
 && C_{n_1, n_2}(p_1 , p_2) \cr \cr
 && C_{n_1, n_2, n_3}(p_1 , p_2, p_3)\cr \cr
 &&\ldots
 \label{conditional correlators}
 \eea

 The corresponding absolute probabilities  go to
 zero as $\exp{-\Delta_D t}$ but the
 conditional
 probabilities have these factors divided out.

Terminals destroy the conformal  theory that lives on the future
boundary but there is evidence that they leave a well defined
scale-invariant structure \cite{Harlow:2011az}.  Linear combinations
of the correlators corresponding to eigenvectors of the rate
equation \ref{ignore terminal} have definite scaling dimension, but
the bigger conformal symmetry is broken. The arguments of
\cite{Harlow:2011az} suggest that two-point correlators of these
scaling fields have the form,%
 \be%
 \langle \Phi_I(Y_1) \Phi_J
(Y_2)\rangle =\sum_n e^{-\frac{S_n}{2}}\Phi_D(n)\Phi_I(n)\Phi_J(n)
|Y_1-Y_2|^{\Delta_D -\Delta_I   -\Delta_J}.
 \ee%

The fact that these are not diagonal in dimension shows that the
special conformal transformations that hold two points fixed on
future infinity are no longer symmetries. The symmetries that do
remain (the symmetries of the conditional correlations) are  scale
transformations, translations, and rotations of the future boundary.

In the absence of terminals there is no preferred frame of reference. The
same symmetry that acts as conformal symmetry on future infinity acts in the bulk
on the time coordinate, transforming from one frame to another. The breaking of conformal symmetry by terminals picks a preferred reference frame or time coordinate, namely the time coordinate for which the correlation functions tend to zero exponentially. It may seem surprising that terminals prejudice the choice of reference frame, but this is not a new effect. In \cite{Garriga:2006hw} it was called \it The Persistence of Memory. \rm The point of that paper was that one must start with an initial condition such that on some space-like surface no terminal bubbles are present. The memory of that surface is not erased by time if there are terminal vacua.

However, apart from the choice of reference frame, correlation functions in \cite{Harlow:2011az} were shown to be universal, scale invariant, and controlled by the dominant eigenvector of the rate equations.

These are the lessons that we will apply to \d2.

\subsection{The Fixed Point}\label{confsect}
All of the non-perturbative effects described in section \ref{NPsect} except for collisions of $\Lambda=0$ bubbles on the same moduli space arose from non-perturbative fluctuations in region II.  As we discussed in section \ref{ds2sect}, at least perturbatively region II has \d2 symmetry.  If we for a moment neglect these same-moduli space collisions, it seems reasonable to apply the lessons of the last section to these non-perturbative processes in \d2. dS bubbles, non-BPS AdS bubbles, and bubbles of disconnected moduli space all produce circular regions on the census taker's sky which contain smaller fractal structure inside of them, and we will argue that they can be thought of as ``dS bubbles'' in the \d2 theory.  BPS domain walls with crunching AdS regions on the other side produce ``dead'' regions, which contain no smaller structure.  We will claim that these can be thought of as terminals in the \d2 theory.  For clarity we will postpone further discussion of these BPS domain walls to the following section.  We will also continue to ignore same-moduli space collisions for a little longer.

\begin{figure}[h]
\begin{center}
\includegraphics[scale=.4]{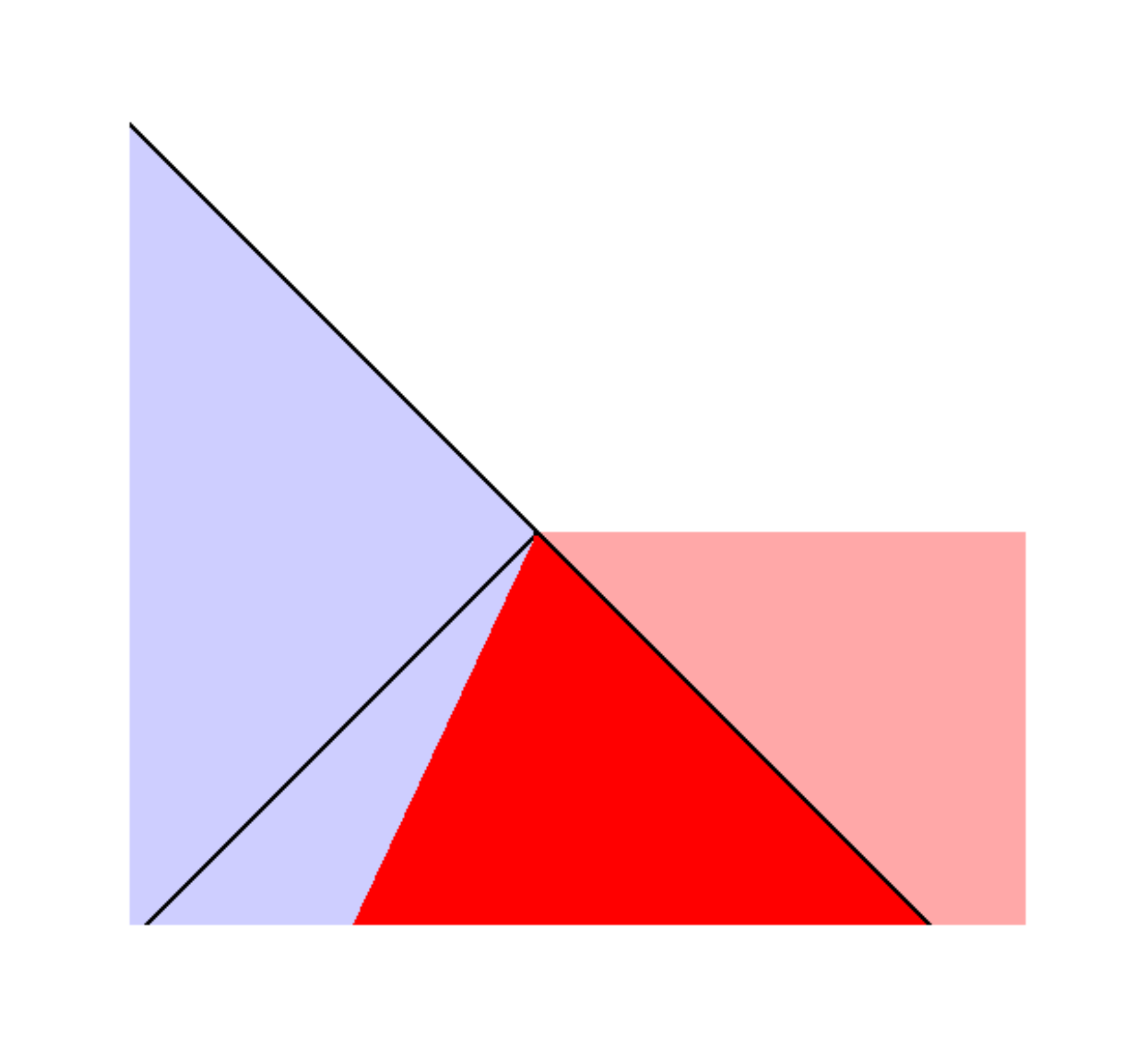}
\caption{A FRW bubble in a red ancestor vacuum. }
\label{fig:10}
\end{center}
\end{figure}

Begin with an FRW bubble in a red ancestor vacuum as shown in Figure \ref{fig:10}.         
From the \21 \ perspective this is just a red de Sitter space. Most of the degrees of freedom are located near $X=0$ (see section 4, equation \ref{entropy of d2}). To keep track of the vacuum adjacent to the FRW region we can imagine painting $\Sigma$ red.

\begin{figure}[h]
\begin{center}
\includegraphics[scale=.4]{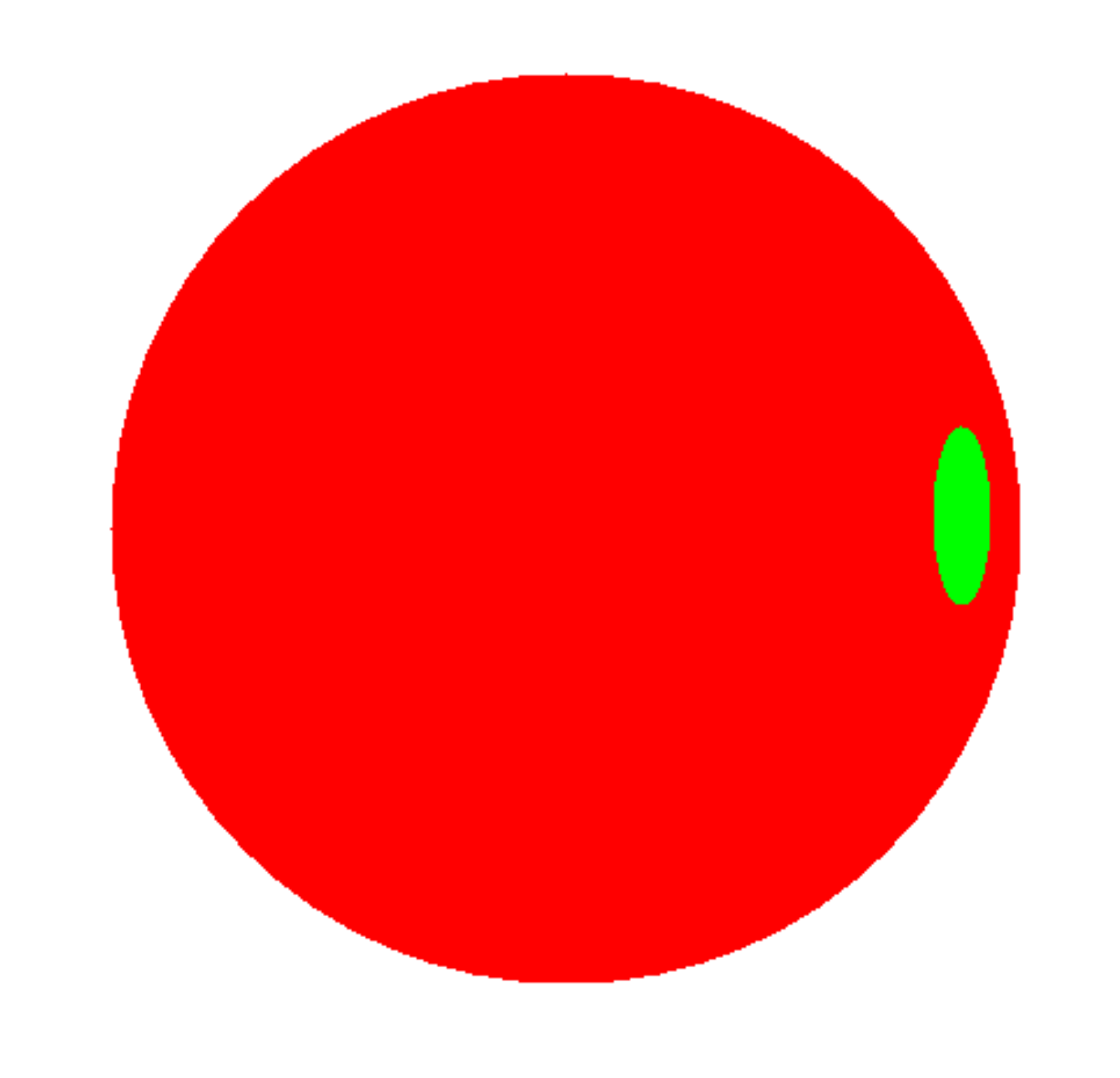}
\caption{A de Sitter Bubble in region II blocks out a disc in $\Sigma.$ }
\label{fig:8}
\end{center}
\end{figure}

Next consider the  nucleation of a de
Sitter bubble of  vacuum type green,  taking place in region
II.\footnote{This argument also applies to AdS bubbles which have non-BPS domain walls with the census taker bubble, as well as $\Lambda=0$ bubbles on a disconnected moduli space.  For simplicity we will continue to refer to them just as dS bubbles.}
\begin{figure}[h]
\begin{center}
\includegraphics[scale=.4]{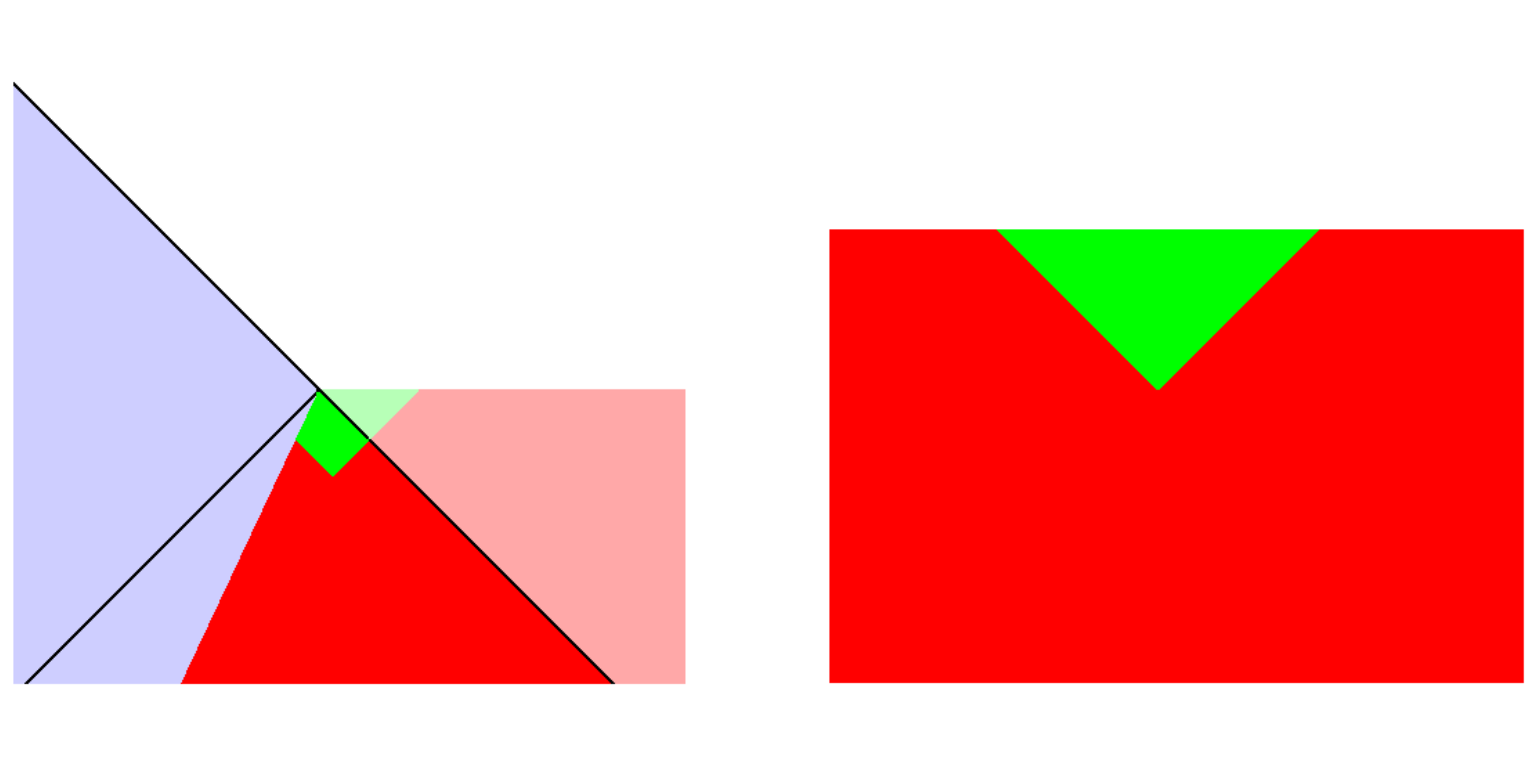}
\caption{Collision with a green de Sitter bubble in the 3+1 geometry (left) appears as a nucleation event in the 2+1 geometry (right). }
\label{fig:11}
\end{center}
\end{figure}
The green bubble will collide
with with the original FRW bubble as shown in the left panel of Figure \ref{fig:11}. From the \21 \ viewpoint this is a bubble nucleation of a conventional kind, as shown in the right panel of Figure \ref{fig:11}.
The bubble will grow and eventually occupy a disc-like region on $\Sigma.$ We will imagine the disc to be painted green as in Figure \ref{fig:8}

Strictly speaking, a nucleation event is
characterized by more than just the vacuum type. In addition to
knowing when the nucleation event took place ($\omega$) we also have
to know where along the $X$ axis it occurred. We may try to keep track of the location of the nucleation event by including an additional label. However, this seems awkward. One should keep in mind that a de Sitter vacuum has many hidden labels associated with the large number of microstates, all of which are lumped into a single macroscopic state. The label $X$ is just another one of these degrees of freedom. Thus
it is  convenient to
integrate over the $X$ coordinate of the nucleation site. The following facts suggests that we integrate over $X,$ keeping  $\omega$ fixed.

\bi
\item A natural definition of the RG flow parameter describing $\Sigma$ is the  coordinate size of a perturbation on the two-sphere $\Omega_2.$ Two bubbles of the same coordinate size and identical color will have similar appearance to the census taker. So it makes sense to lump then together.

     \item The coordinate size of a green disc on $\Sigma$ is a function of where, in region II, the green bubble nucleated. In other words it is a function of $\omega $ and $X.$ However, when the geometry of bubble collisions is examined, it is found that the coordinate size of the green disc only depends of $\omega.$ It is independent of $X.$ This is shown in the appendix.

\ei

 From the census taker's viewpoint there is very little difference between bubbles that nucleate at $(\omega, X)$ and $(\omega, X').$ We can absorb the $X$ coordinate into the internal degrees of freedom of the bubble by integrating over $X.$ This allows us to organize bubbles according to a natural scale-size.
 Once that is done, bubble nucleation in region II can be identified with \21 \ bubble nucleation in \d2, and the arguments of \cite{Harlow:2011az} should apply. In particular the analysis of Section 9 based on rate equations should hold. Note that the time parameter in the rate equations is replaced by $\omega.$
For example, the fractional coordinate area $P(n)$ satisfies the rate equation
\be
  \frac{dP_n}{d\omega} =     \dot{P}_n=\sum_m (\gamma_{nm}P_m -\gamma_{mn}P_n)
  \label{rate-omega}
\ee

The rest of the analysis is  the same as in
Section 9. In particular, in the absence of terminal \21 \ vacua, there is a stationary probability
distribution representing a universal bath of
 bubbles, contiguous with $\Sigma.$ As the cutoff ($\omega$) tends to
 infinity the distribution  becomes independent of the initial conditions and conformally invariant.

The discretuum of
 very low dimensional conformal \it multiverse fields \rm that were described in \cite{Harlow:2011az} will be present in the spectrum of dimensions in the \21 \ dimensional theory.

\begin{figure}[h]
\begin{center}
\includegraphics[scale=1]{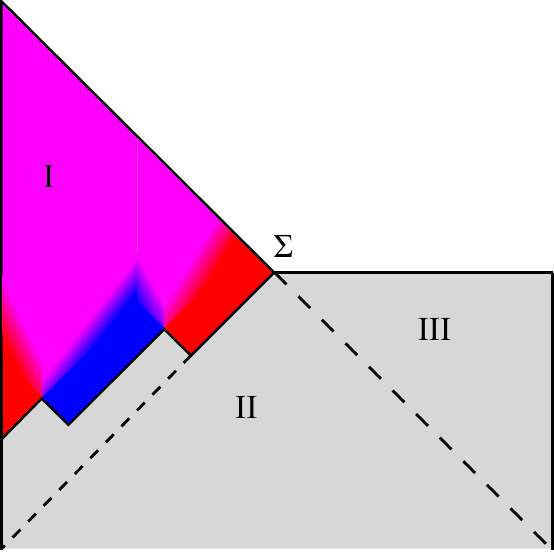}
\caption{Same-moduli space collisions as region I dynamics.}
\label{reg1dyn}
\end{center}
\end{figure}

Let's now consider collisions with bubbles on the same moduli space.  As we saw in \ref{fig:5}, these collisions simply merge with the hat and push it out a little bit.  They can really be thought as part of the dynamics of region I.  As more bubbles are nucleated at smaller scales and in different directions a fractal pattern of values of the moduli will be produced throughout region I.  It is this fractal distribution that we truly identify as hat, and it is clear that for a given connected moduli space there can only be one type.  In figure \ref{reg1dyn} this procdure is shown for a few collisions.  It may appear that the bubble that the census taker enters first is special, but from \ref{fig:5} it should be clear that there is symmetry between these different colliding bubbles and the final pattern cannot depend on this.  This distribution is the window through which the census taker looks down on region II.

An important issue is that we don't actually expect that the region I dynamics are decoupled from those in region II, as our discussion might have implied.  Both of these dynamics affect the generation of the pattern on the census-takers sky, and there will be some coupling between them.  The real proposal is that there is some big rate equation that involves both the region I dynamics and the region II dynamics, and that together it flows to a joint universal fixed point.

Whether or not the fixed point is the same for hats that correspond to \textit{different} moduli spaces is a more difficult question.  We see no reason why hats with disconnected moduli spaces need to be described by the same fixed point.  A more subtle issue is when there are different moduli spaces that are connected at infinity, as often happens in string theory where different moduli spaces are usually (always?) connected by a decompactification limit.  Perhaps this connection is enough to combine the moduli spaces into a single type of hat, but more analysis is clearly required and we leave it for future work.

To summarize, we expect the theory on $\Sigma $ to be characterized by three important features:

 \bi
 \item The fixed point behavior of the FRW/CFT is universal. It does not
 depend on any initial condition including the ancestor vacuum.  If there are disconnected moduli spaces then it seems likely to depend on which one the hat ends up in, so we probably need to condition on this.

 \item The landscape of de Sitter vacua is reflected in a discretuum of very
 low dimension multiverse operators.

 \item  Unlike the global dS/CFT the entire field theory on $\Sigma$ is
 visible to a single observer. That observer does not live on \d2, but rather
 in the hat. Unlike the perturbative case, the radiation entering the hat
 from the initial condition surface carries information about the entire
 multiverse.

 \ei

The discretuum of low dimension operators is unusual but it has an important meaning that deserves discussion.
Operators of low dimension are relevant in the RG sense and
 imply  that the fixed point is unstable as the flow proceeds toward the
 infrared. A large number of such relevant operators indicates that the  flow
 may split up into a equally  large number of distinct IR limits.
One can turn it around and say that a great many starting points in the IR
will all lead to the same UV fixed point.  These IR starting points simply
reflect the multiplicity of the landscape itself. The discretuum of de Sitter vacua is directly connected to the discretuum of low dimensional fields on $\Sigma.$ It is possible that this provides a precise characterization of the Landscape of de Sitter vacua.

These low dimension operators also play an important role in determining leading corrections to correlators as the UV fixed point is approached.  To understand this consider a general $d$ dimensional CFT  in finite volume ($\sim L^d$) perturbed by a relevant operator $O_j$ with dimension $\Delta_j$.  The action is

\be
S = S_{\rm CFT} +u_j\int d^d z~ O_j(z)
\ee

The strength of the perturbation is set by demanding that its effect be at most order one at the infrared scale, where the FRW/CFT flow to the ultraviolet  begins.  So $u_j  \sim (1/L)^{d-\Delta_j}$. The two point function of another operator $O_i$ can be computed perturbatively in $u_i$.

\be
\langle  O_i(x) O_i(y) \rangle = \langle O_i(x) O_i(y)\rangle + u_j \int d^d z ~\langle O_i(x) O_i(y) O_j(z) \rangle + \ldots
\ee

Using the conformally invariant form for the three point function and assuming that the operator product coefficient $C_{iij}$ is nonvanishing we find

\be
\langle  O_i(x) O_i(y) \rangle = \frac{1}{|x-y|^{2 \Delta_i}} + u_i \int d^dz ~ \frac{C_{iij}}{|x-y|^{2 \Delta_i- \Delta_j}|x-z|^{\Delta_j}|y-z|^{\Delta_j}}+ \ldots
\label{xx}
\ee

There are two types of operators $O_j$ that produce slow transients as the UV fixed point is approached.  The first are almost marginal operators, $\Delta_j \sim d$.  The intuition here is clear.  Under the renormalization group flow it takes a long time for the perturbation to the action to fade away.  This can be seen by evaluating the integral in \ref{xx}.  It is UV and IR convergent and so its value can be determined by scaling.  Absorbing a constant we find
\be
\langle  O_i(x) O_i(y) \rangle = \frac{1}{|x-y|^{2 \Delta_i}}\left(1 + u_i  |x-y|^{d-\Delta_j}\right) +\ldots
\label{yy}
\ee

The UV fixed point can be seen in correlators where $|x-y| \ll L$.    The approach to the fixed point is determined by the correction term in \ref{yy} which is
$ \sim |\frac{x-y}{L}|^{d-\Delta_j}$ where the characteristic size of $u_j$ has been inserted.

A second source of slow transients are highly relevant low dimension operators of the kind discussed above.   Here $\Delta_j \ll d$.   At first glance it is surprising that these would give rise to slow transients.  Their strength in the action rapidly decreases in the flow to the UV.   But their effect on correlation functions is enhanced by IR divergences.  In particular for $\Delta_j < d/2$ the integral in \ref{xx} is IR divergent.  Keeping the leading IR divergent term in \ref{xx} we obtain

\be
\langle  O_i(x) O_i(y) \rangle = \frac{1}{|x-y|^{2 \Delta_i}}\left(1 + u_i  |x-y|^{\Delta_j}L^{d-2\Delta_j}\right) +\ldots
\label{zz}
\ee
Inserting the characteristic size of $u_j$ we find a slow transient  of order $  \sim |\frac{x-y}{L}|^{\Delta_j}$ .
The slow transients in the tree model produced by the low dimension multiverse fields are of this type.   Higher order terms in $u_j$ do not alter these qualitative behaviors.

Leading order perturbative correlations in the doubled dS/CFT contain conjugate pairs of operators with dimensions $\Delta$ and $d-\Delta$.   For small mass bulk scalars one of these is highly relevant, the other almost marginal.  The above analysis shows that they contribute the same type of slow transient in the flow to the UV.  At higher orders in bulk perturbation theory the precise pairing is broken \cite{Marolf:2010zp, Jatkar:2011ju} .    The multiverse fields are highly nongaussian \cite{Harlow:2011az}   but for low nucleation rates it seems plausible that the full doubled theory will contain nearly marginal as well as highly relevant operators.

As mentioned earlier, the non-perturbative effects induce configurations on $\Sigma$ which can reasonably be called  instantons. It is the role of instantons to mix otherwise inaccessible regions of target space. Moreover small instantons can induce non-perturbative RG flows. From the FRW/CFT point of view that is exactly what is happening.

\setcounter{equation}{0}
\section{Terminals and BPS Domain Walls}\label{terminalsect}

There are a number of reasons to believe that de Sitter space cannot be  absolutely stable and must decay to terminal vacua. In particular there is a rigorous  no-go theorem stating that the symmetries of de Sitter space cannot be realized in a system with finite entropy \cite{Goheer:2002vf}. That theorem applies to \d2.

In the usual framework of \31 \ de Sitter space, the way out of the theorem is the existence of terminals, which as shown in \cite{Harlow:2011az} and reviewed in section 9, breaks the de Sitter symmetries. The same mechanism should apply in the lower dimensional case, and indeed we believe it does.

The candidates for \21 \ terminals are BPS domain walls which can form when a supersymmetric ADS vacuum collides with the census taker's bubble \cite{Freivogel Horowitz Shenker}.
The geometry of the BPS domain wall   is flat. In fact, in the lower dimensional sense the BPS domain wall
is a terminal $(2+1)$-dimensional FRW geometry. To put it another way, the
BPS domain walls provide exact $(2+1)$-dimensional analogs of the
$(3+1)$-dimensional  supersymmetric hat-geometries.

\begin{figure}[h]
\begin{center}
\includegraphics[scale=.4]{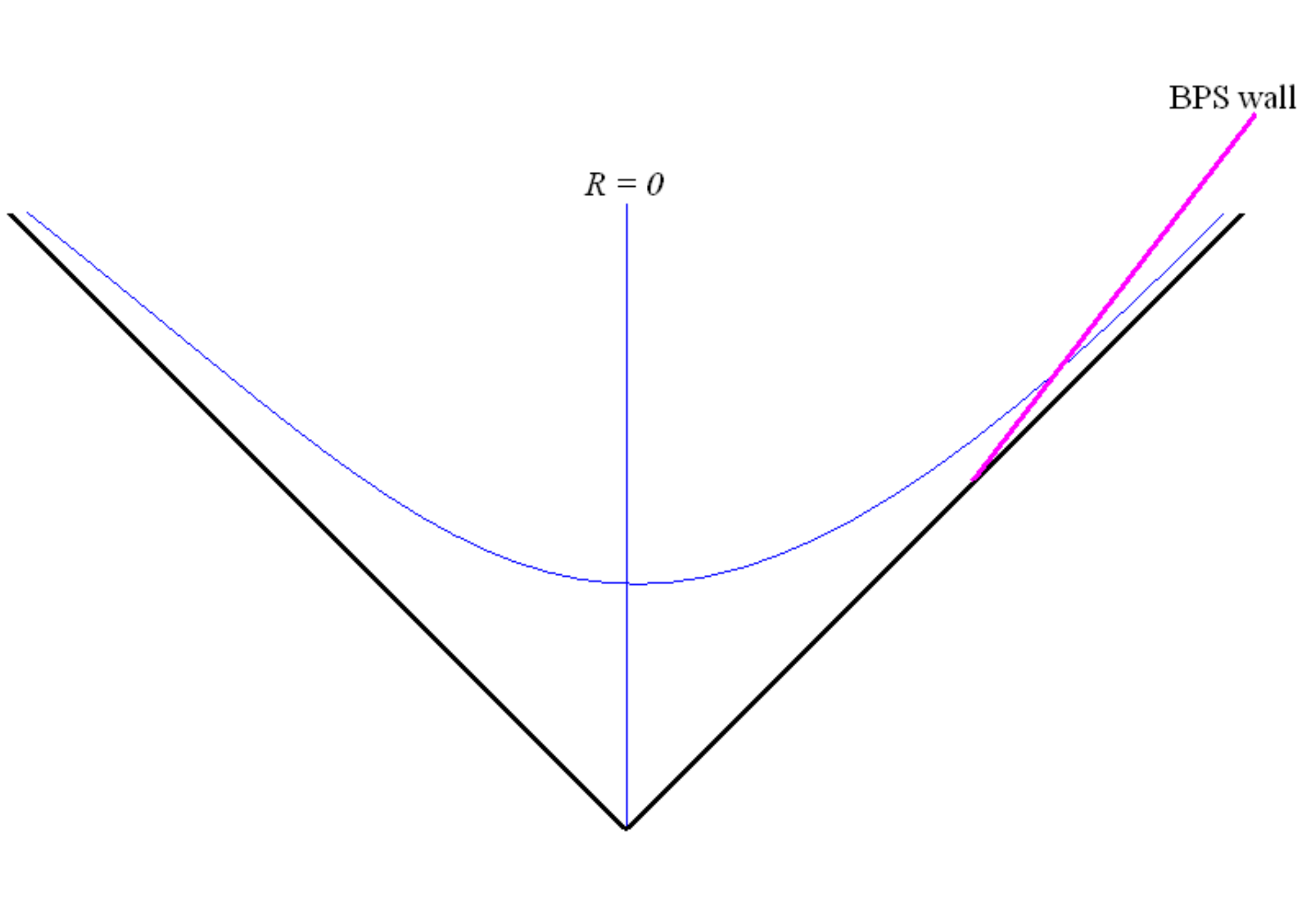}
\caption{A BPS domain wall is a non-accelerating plane in the flat FRW
geometry.  }
\label{fig:9}
\end{center}
\end{figure}

The BPS domain walls do not destroy the \31 \ census taker. In this respect
the conformal diagram Figure \ref{fig:7} \ is misleading. Instead we can
consider a metrically more faithful version in which at late-time, region I
becomes the interior of a Minkowski-space future light cone. The BPS domain
walls are non-accelerating planes that typically never get very deep into the
region near $R=0.$ This is shown in Figure \ref{fig:9}.

Their effect is precisely the lower dimensional analog of the effects discussed in
\cite{Harlow:2011az} and Section 9. The rate equation \ref{rate-omega} gets an additional term as in \ref{ignore terminal},
\be
  \frac{dP_n}{d\omega} =     \sum_m (\gamma_{nm}P_m -\gamma_{mn}P_n)-\gamma_n
  P_n
  \label{ignore terminal omega}
\ee %

Again the leading time dependence is governed by a dominant eigenvector with eigenvalue
$e^{-\Delta_D}$ ($\Delta_D >0$) and the fractional coordinate area occupied by non-terminals goes to zero as $e^{-\Delta_D\omega}.$ However the proper area of the non-terminal region increases as $e^{(2-\Delta_D)\omega}.$

Once there are terminals it becomes important to specify an initial condition
on some spacelike hypersurface in the multiverse. We will follow Garriga,
Guth, and Vilenkin \cite{Garriga:2006hw} and define the initial condition by
specifying an initial surface with a given color and no terminal bubbles present. The particular
hypersurface breaks the symmetry of de Sitter space  by  singling out a
preferred frame, namely the frame of the initial surface.
 If terminal vacua did not exist, that symmetry breaking would heal itself
 with time, but in the presence of terminals the symmetry breaking persists.

Conditional  questions, of the kind described by \ref{conditional correlators}, concerning the
inflating fractal can be asked.
The eigenvalues of \ref{ignore terminal} are all less than $1.$ The stationary
eigenvector is  replaced by a \it dominant \rm eigenvector with a very small
eigenvalue $e^{-\Delta_D.}$ In addition there remains a discretuum of very
small sub-leading scaling dimensions $\Delta_I,$  all satisfying $1>> \Delta_I >
\Delta_D.$

As in Section 9, terminals destroy the conformal field theory that lives on $\Sigma,$ but leave  well defined conditional correlation functions.
Given two points $\sigma_1$ and $\sigma_2,$  the conditional
probability that they have color $n_1$ and $n_2,$ given that neither has been
swallowed up by a terminal bubble, is defined to be
\be
C_{n_1n_2}(Y_1 , Y_2).
\label{CnY}
\ee
The conditional correlations are scale invariant but not conformally invariant. In particular the two point function \ref{CnY} scales but is non-diagonal in the dimensions of the operators. This shows
 that the special
conformal transformations that hold two points fixed on $\Sigma$ are no
longer symmetries. The symmetries that do remain (the symmetries of the
conditional correlations) when BPS terminals are included are,
\bi
\item Scale Transformations $Y^i \to \lambda Y^i.$
\item Rotations of the $Y$ plane.
\item Translations of the $Y$ plane.
\ei

It is clear that the effect of terminals is to change the attractor behavior to something new. In \cite{Harlow:2011az} we called it a \it  fractal flow. \rm The fractal flow is a universal attractor with correlation functions that scale, but it is not a field theory in any ordinary sense. Among other things, it lacks standard cluster-decomposition properties. It does have one remarkable property that we will come back to: unlike an conformal fixed point, the fractal flow leads to an arrow-of-time. In the present context it means that the lower-dimensional $2+1$-dimensional geometry also has an arrow-of-time. 

The breaking of conformal invariance by BPS domain walls is perhaps most obvious
from the \d2 \ description of region II but it also has implications for the FRW/CFT duality in region I.
The breakdown of symmetry is a form of
persistence of memory described in \cite{Garriga:2006hw}. There are two time-scales that
relate to the persistence of memory. Consider the global initial condition for the multiverse,  defined as a surface on which there are no terminal bubbles. By convention let the time of that surface be $t=0.$  The census taker's bubble nucleates at a later time, $t_n.$ Inside the Census taker's bubble, the FRW time from the nucleation event can be called $t.$  The persistence of memory states that for fixed $t,$ there is a breaking of spatial homogeneity ($O(3,1)$ symmetry) in the FRW patch that does not disappear as $t_n \to \infty.$ This effect is a consequence of terminal vacua in the ordinary \31 \ sense. It does not in itself imply that the violation of spatial homogeneity persists to arbitrarily large FRW time $t\to \infty$ within the census taker's bubble. If the theory on $\Sigma$ runs to a conformally invariant fixed point then the $O(3,1)$ symmetry will be restored.

On the other hand the added ingredient of BPS domain walls breaks the $O(3,1)$ of $\Sigma.$ This is a stronger persistence that implies that the symmetry between census takers in the FRW patch is never restored as $t\to \infty.$
 There is a preferred census taker who can claim to be at the center of the spatial geometry.

The census taker at $R=0$ may be special, but he is in the minority. All but
a negligible fraction of census takers are out at asymptotic $R.$ Therefore
the limiting  coordinates $U^{\pm}, Y^i$ are especially appropriate for
polling the census takers in a given FRW cosmology. There left-over symmetry
is sufficient to make sure that asymptotic Census  Takers at large $R$ are
all equivalent.

Finally it  is obvious that the radiation entering the hat will be affected
by terminal  bubble nucleation, replacing the fixed point with something a
little different.

\setcounter{equation}{0}
\section{Region III}

In this paper we have had almost nothing to say about region III other than that it exists and borders on $\Sigma.$  The most interesting questions are about region III and its relation to the census taker. What is clear is that just as the surface $\bf{a}\it$  defines initial conditions for the FRW region, the surface $\bf{b}\it$  defines initial conditions for the rest of the multiverse: region III. Obviously the phenomena that take place in region II create initial conditions for both I and III.

From the point of view of the census taker region III is analogous to the region behind the horizon with the census taker playing the role  of an observer who remains outside the black hole.  As in the black hole case, the census taker receives Hawking radiation from the horizon, i.e., the surface that we earlier called $\bf{b}\it.$  Unlike the black hole case, the amount of radiation is infinite, which raises the question of whether it can encode the entire multiverse. Obviously we think the answer may be yes. One way of expressing the issue is to ask if the Hawking radiation is complementary to information in region III, and if so, can complementarity provide a tool for addressing the measure problem?

\begin{figure}[h]
\begin{center}
\includegraphics[scale=.4]{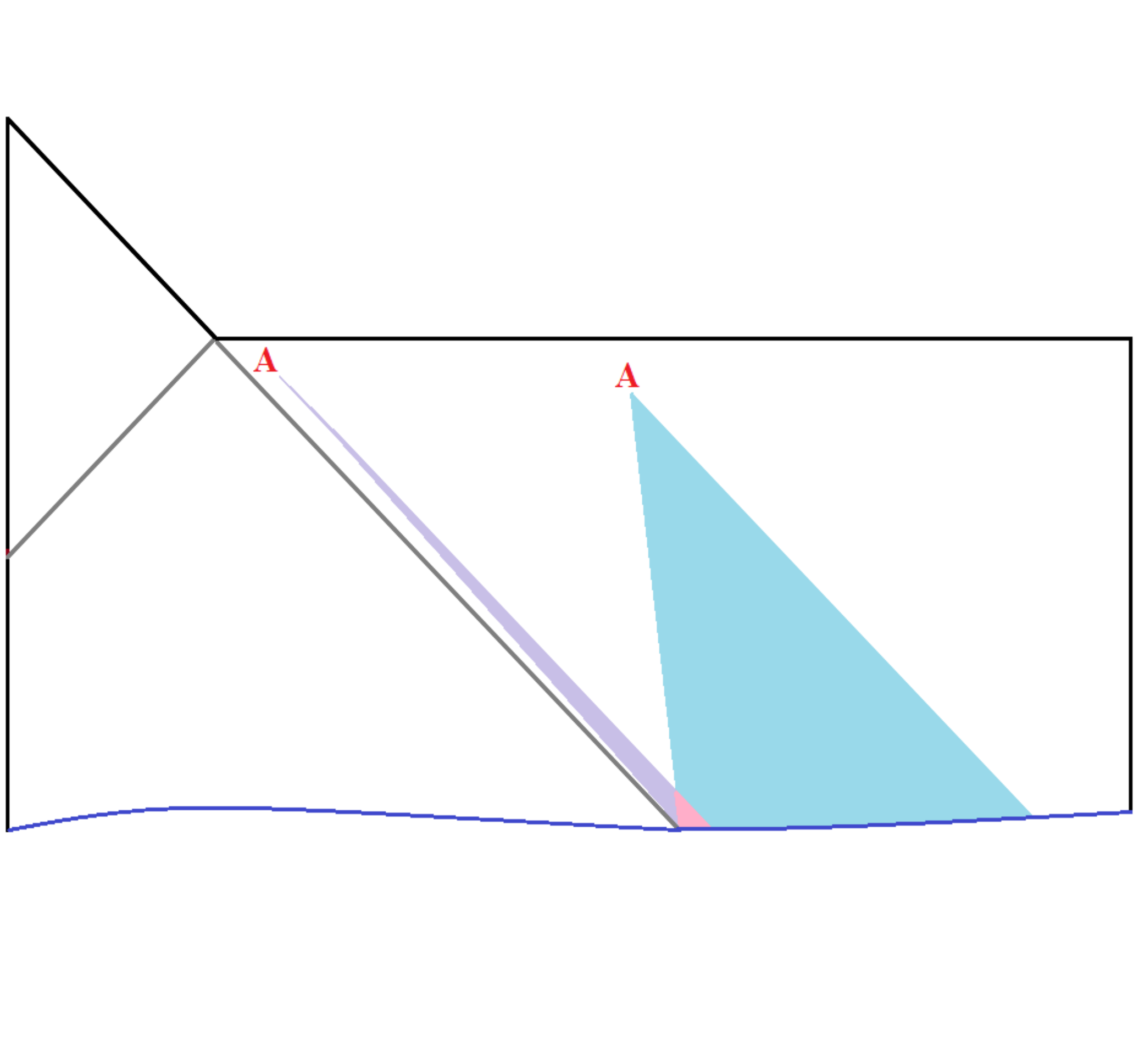}
\caption{As the story $A$ is moved toward $\Sigma$  all but a vanishingly small fraction of its
information passed through the horizon.  The dark blue line at the bottom of the figure represents the
global initial condition.}
\label{fig:zero}
\end{center}
\end{figure}

Complementarity requires that the information in region III can be ``pulled back" through the horizon to region II, and then run forward into region II. It is not obvious that all information in III can be pulled back in this way. Just to illustrate, consider figure \ref{fig:zero}. For a story $A$  located in a generic point in region III, the portion of its causal past represented by light blue, does not intersect the census taker's horizon. However, it is also clear from the figure, that as $A$ is moved closer to $\Sigma,$ the missing fraction of information tends to zero. That is one of the reasons for zooming in on $\Sigma$ as we did in Section 1.
In that limit no information gets to $A$ without passing through the horizon. Thus, in the limit,  it should be possible to pull every story back into region II. Since the entire multiverse is replicated in every small comoving volume, zooming in should not be a real limitation in studying the rest-of-the-multiverse.

The problem of how horizons encode, or perhaps encrypt is a better word, information behind horizons is a new kind of quantum-information-theoretic question which is receiving recent attention \cite{Hayden:2007cs}  \cite{Sekino:2008he} \cite{Lashkari:2011yi}.

\section{Conclusions}
Let us list and  review  the conclusions of this investigation:

\bi
\item The first point---one which we have emphasized in earlier publications---is that a precise formulation of quantum-gravitational systems requires  the existence of a causally connected region which has access to an infinite amount of information \cite{Harlow:2010my}. This precludes conventional closed FRW cosmologies, and stable eternal de Sitter vacua. If one adds the condition \cite{Bousso:2011up} that irreversible de-coherence can occur, that leaves eternal inflation as the only candidate. The causally connected regions with infinite information are the supersymmetric FRW geometries bounded by hats. The holographic description of such a region is the FRW/CFT theory located on the boundary-surface $\Sigma.$

\item The surface $\Sigma$ is more that just the location of the  FRW/CFT holographic duality. It is the intersection of three regions. Region I is the census taker's FRW region  bounded by a hat. Region II is a $(2+1)$-dimensional de Sitter space called \d2.  Region III is the rest of the multiverse behind the census taker's horizon.
\item  Unlike the usual global \31 \ de Sitter space, the entire future boundary of \d2 is can be seen by a single observer. In that sense the theory on $\Sigma$ has operational significance that a theory on the future boundary of \31 \ de Sitter space would not have. The census taker plays the role of a meta-observer for \d2.

\item While \d2 has a finite de Sitter entropy, the total amount of information visible to the census taker is infinite. The meaning of this unbounded amount of information is not certain, but we conjecture that it is complementary to the unbounded information in region III---the rest of the multiverse.

\item In perturbation theory, the FRW/CFT fixed point on \sig \ is specific to a particular \cdl \ decay.  Howerver, non-perturbative effects of bubble collisions lead to a mixing of the initial vacua and the creation of a more universal fixed point that does not depend on the initial ancestor vacuum.

 \item \d2 \ is a genuine quantum de Sitter space with a dynamical gravitational field. It defines a lower dimensional version of eternal inflation. In the absence of terminal vacua  the  2-dimensional dS/CFT theory living on \sig \ is conformally invariant.

\item BPS domain walls connecting the census taker's vacuum with supersymmetric ADS bubbles correspond to terminal vacua in the \21 \ eternally inflating theory in \d2. Such terminals break the conformal invariance on \sig, which in turn means that spatial homogeneity is broken in the census taker's FRW patch. This is a strong form of persistence of memory and lead to a preferred spatial center in FRW, for all time.

\item  The existence of large number of \31 \ de Sitter vacua leads to an equally large number of very low dimensional multiverse fields. These multiverse fields are very relevant in the RG sense. Therefore the flow to the infrared can be expected to break up into a huge number of branches. These branches correspond to different initial starting vacua for eternal inflation. This may provide a way of classifying vacua by operators in the FRW/CFT theory.

\ei

\section*{Acknowledgements}
Some of the ideas presented  in this paper build on earlier unpublished work on the ``Census-Taker Measure" done in collaboration with Raphael Bousso, Ben Freivogel, Alex Maloney,  and I-Sheng Yang.  We are grateful to them for their insights.

Our work is supported by the Stanford Institute for Theoretical Physics and NSF Grant 0756174. DS also acknowledges the NSF under the GRF program.

\setcounter{equation}{0}
\section{Appendix}

\begin{figure}[t]
\begin{center}
\includegraphics[scale=1]{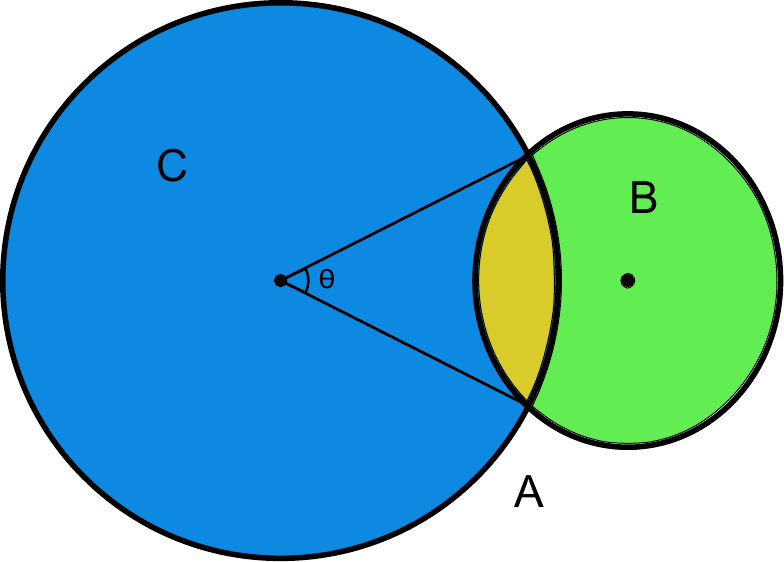}
\caption{The region subtended at future infinity by a collision of a census-taker bubble $C$ with a dS bubble $B$ inside of an ancestor vacuum $A$.  What happens inside the yellow region depends on the details of the domain wall between $B$ and $C$, but it does not affect the angle the collision influences on the census-taker's sky, which is denoted $\theta$.}
\label{collision}
\end{center}
\end{figure}

In this appendix we derive the relationship between the nucleation time of a bubble of some dS vacuum B in region II and its size on the census-taker's late-time sky.  The metric of the ancestor dS space in flat slicing is
\begin{equation}\label{appflat}
ds^2=\frac{1}{T^2}\left(-dT^2+d\vec{x}^2\right).
\end{equation}
We can take the census taker's bubble to nucleate at $\vec{x}=0$, $T=-1$, and we will take the bubble of vacuum $B$ to nucleate at $\vec{x}=\vec{x}_B$, $T=T_B$.  These nucleations will remove the union of two spheres from future infinity in the ancestor dS space, one centered at $\vec{x}=0$ with radius one and the other centered at $\vec{x}=\vec{x}_B$ with radius $-T_B$.  The affected angular size on the census-taker's late-time sky is shown in figure \ref{collision} and is given by
\begin{equation}\label{ctangle}
\cos\left(\frac{\theta}{2}\right)=\frac{1-\vec{x}_B^2+T_B^2}{2|\vec{x}_B|}.
\end{equation}

We'd like to recast this in terms dS slicing coordinates in region II, in terms of which the metric is
\begin{equation}\label{appdS}
ds^2=d\xi^2+\sin^2 \xi \left[-d\omega^2+\cosh^2 \omega d\Omega_{d-1}^2\right].
\end{equation}
By looking at the relationship of the coordinates \eqref{appflat} and \eqref{appdS} to the embedding coordinates in $d+2$ dimensional Minkowski space we find
\begin{align}
\nonumber
&\sin \xi \sinh \omega=\frac{1}{2}\left(T-\frac{1+\vec{x}^2}{T}\right)\\
&\sin \xi \cosh \omega = -\frac{|\vec{x}|}{T},
\end{align}
whose ratio then gives
$$\tanh \omega=\frac{1+\vec{x}^2-T^2}{2|\vec{x}|}.$$
From \eqref{ctangle} we thus see that the angle subtended on the census-taker's sky is simply related to the time in the dS slicing via
$$\cos \left(\frac{\theta}{2}\right)=\tanh \omega_B.$$
Interestingly the angle is independent of the spatial coordinate $\xi$, so the dS slicing is particularly natural for talking about how nucleation events in region II affect the census-taker's observations.

\end{document}